%% file: m-t.tex
\documentclass{aa}  
\usepackage{times,epsfig,amssymb,amsmath}

\newcommand{\chandra}{{\it Chandra} }  
\newcommand{\asca}{{\it ASCA} }  
\newcommand{\sax}{{\it BeppoSAX} }  
\newcommand{\rosat}{{\it ROSAT} }  
\newcommand{\xmm}{{\it XMM-Newton}}  
  
\title{Gravitating mass profiles of nearby galaxy clusters and relations  
with X-ray gas temperature, luminosity and mass}  
  
\author{S. Ettori \inst{1} \and S. De Grandi \inst{2}   
\and S. Molendi \inst{3} }   
\institute{  
 European Southern Observatory, Karl-Schwarzschild-Str. 2, D-85748 Garching, Germany     
\and  
 Osservatorio Astronomico di Brera, Via Bianchi 46, I-23807 Merate  
(LC), Italy   
\and  
 Istituto di Fisica Cosmica ``G.Occhialini'', Via Bassini 15, I-20133  
Milano, Italy   
}  
  
\offprints{S. Ettori}  
\mail{settori@eso.org}  
  
\date{Received 24 February 2002 / Accepted 7 June 2002}  
  
\begin{document}  
\titlerunning{Cluster mass profiles using \sax}  
  
\abstract{   
We consider a sample of 22 nearby clusters of galaxies observed with the  
Medium Energy Concentrator Spectrometer (MECS) on board 
\sax. They cover the range in gas temperature between 3 and 10 keV,  
with bolometric X-ray luminosity between $2 \times 10^{44}$ erg s$^{-1}$  
and $6 \times 10^{45}$ erg s$^{-1}$.   
Using the de-projected gas temperature and density profiles 
resolved in a number of bins between 5 and 7 and obtained 
from this dataset only, we recover the total gravitating mass profiles 
for 20 objects just applying the (i) spherical symmetry 
and (ii) hydrostatic equilibrium assumptions.   
We investigate the correlations between total mass, gas temperature  
and luminosity at several overdensities values and find that the slopes  
of these relations are independent of the considered overdensity  
and consistent with what is predicted from the cluster scaling laws.  
The best-fit results on the normalization of the $M-T$ relation    
are slightly lower, but still consistent considering
the large errors that we measure, with hydrodynamical
simulations.
A segregation between relaxed and non-relaxed systems is 
present in each plane of these relations pointing out 
a significant component in their intrinsic scatter.  
This segregation becomes more evident at higher overdensities 
and when physical quantities, like $M_{\rm gas}$ and $L$, that 
are direct functions of the amount of gas observed, are considered. 
}  
  
\maketitle  
  
\section{Introduction}   
   
The amplitude and the shape of the power spectrum of the primordial density  
fluctuations on scales of about 20 $h_{50}^{-1}$ Mpc can be effectively  
constrained with the mass function of galaxy clusters.  
Since the early '90s, X-ray observations have been used 
to build large datasets  
of measured luminosities and, with more effort because a larger   
number of source counts is required, temperatures of the   
X-ray emitting plasma trapped in the cluster gravitational potential.  
These observed quantities are expressions of the physical processes that  
are taking place in the galaxy clusters and manifest the energy and  
the mass of these systems.   
Then, comparing the observed distribution in luminosity (or temperature)  
with theoretical models of the expected cluster number density  
that are functions of total mass and redshift and depend upon the  
cosmological model adopted (e.g. Press \& Schechter 1974),  
it has been possible to put constraints in the ``normalization--shape''   
plane of the primordial density fluctuations spectrum  
(see, e.g., the pioneering work of Henry \& Arnaud 1991 
and the most recent results in Ikebe et al. 2002 and references therein).
  
However, the conclusions reached making this comparison   
rely on an efficient way to relate   
the observed quantities (like gas luminosity and temperature) to the  
gravitating mass of the systems.   
Gas--dynamics simulations (e.g. Evrard, Metzler \& Navarro 1996, Schindler 1996)  
have confirmed the expected correlation between mass and temperature   
and have shown that mass estimates are reliable when obtained  
through X-ray analysis under the assumption of spherical symmetry and  
hydrostatic equilibrium.   
More recently, mass profiles obtained relaxing the condition of plasma  
isothermality have shown a significant mismatch in normalization and slope  
of the mass--temperature relation between observational data and simulations  
(e.g. Horner, Mushotzky \& Scharf  1999, Nevalainen, Markevitch \& Forman 2000).  
On the other hand, it has been clear since the first compilation  
of catalogues of luminosity and temperature (Mushotzky 1984,  
Edge \& Stewart 1991) that the observed correlation between these two  
quantities deviates significantly from the expected scaling law, suggesting  
contributions to the total energy  of the plasma from  physical  
phenomena other than the gravitational collapse.  
  
The observed Luminosity-Temperature ($L-T$) and Mass-Temperature ($M-T$)   
relations for galaxy clusters are, therefore, the foundation   
to construct the cluster mass function and to use these virialized objects
as cosmological probes.  
In this paper, we investigate these relations and, more in general,   
any correlation between observed and inferred quantities using   
\sax observations of 22 nearby clusters of galaxies with resolved   
gas temperature and density profiles.  
The main differences between this study and previous work on the   
same subject are:   
\begin{enumerate}  
\item the use of \sax data that allows us to extend   
the analysis of spatially-resolved spectra up to 20\arcmin\   
in radius, i.e. $\sim 2.5$ times the most favourable configuration   
with \chandra (Weisskopf et al. 2000)
and to put under control some systematic effects  
(e.g., sharper and more energy-independent Point-Spread-Function   
than the \asca one --Tanaka et al. 1994--, 
more stable and lower background than the one  
observed in \xmm\ --Jansen et al. 2001),  
\item the direct deprojection of the spectral results  
to reconstruct the gas temperature and density profiles   
in a model-independent way.  
\end{enumerate}  
  
The sample presented in this work is, to date, the largest for
which the physical quantities (i.e. gas density, temperature,
luminosity, total mass, etc.) have all been derived simultaneously from 
spatially-resolved spectroscopy of the same dataset.
The difference between this approach and  others which make use
of data coming from different satellites and/or make strong assumptions
on the temperature profiles, such as isothermality, is twofold:
on one side the use of data from different missions and the simplistic
assumptions on the temperature profiles allow to build up samples bigger
than ours, on the other they increase the likelihood of systematic effects
which may in turn affect the relations between the observed quantities.

The paper is organized as follows: in Section~2, we describe the   
\sax MECS observations of the galaxy clusters in our sample and   
the results of the spectral analysis considered in this work;  
the deprojection technique applied to the projected spectral results is  
discussed in Sect.~3;   
the gravitating mass profiles are obtained and compared  
with the optical measurements in Section~4; in Section~5, we study the  
correlation between the total mass, gas temperature,   
gas mass and luminosity;  
we summarize our results and present our conclusions in Sect.~6.   
  
All the errors quoted are at $1 \sigma$ level   
(68.3 per cent level of confidence for one interesting parameter)   
unless otherwise stated. The cosmological parameters   
$H_0 = 50 h^{-1}_{50}$ km s$^{-1}$ Mpc$^{-1}$ and  
$\Omega_{\rm m} = 1 -\Omega_{\Lambda} = 1$ are assumed hereafter.  
  
\section{The sample}  
  
We selected from the \sax SDC archive 
all the on-axis pointings of galaxy clusters with 
redshift smaller than $\sim 0.1$ and
exposure times larger than 30 ksec.
The observation log for the cluster sample, with a detailed discussion  
of the metal abundance and temperature profiles derived for   
subsets of this sample, is given in De Grandi \& Molendi (2001 and
2002, hereafter DGM02).  
  
In this paper we discuss data from the imaging Medium-Energy  
Concentrator Spectrometer (MECS; 2-10 keV; Boella et al. 1997). The  
MECS consists of two identical grazing incidence telescopes with  
imaging gas scintillation proportional counters in their focal planes.  
The field of view of the MECS is circular with radius of $\sim  
25^\prime$. This detector has a spectral resolution of $\sim 8\%$ at 6  
keV and a Point Spread Function (PSF) of $\sim 1^\prime$ (HPR), which  
varies only weakly with the energy (D'Acri, De Grandi, \& Molendi  
1998).  The MECS has an entrance Beryllium window sustained by a  
thicker supporting structure, or strongback, in form of a circular  
ring and four ribs, which has transmission properties different by the  
rest of the window.  
  
The data analysis is fully described in De Grandi \& Molendi (2001,  
2002), hence in this paper we will only summarize the whole procedure.  
Standard reduction procedures and screening criteria have been applied  
using the SAXDAS package under the FTOOLS environment to produce  
equalized and linearized MECS event files.   
Each cluster has been divided into concentric annuli centered on the  
X-ray emission peak 
computed by fitting a Gaussian to the photon
distributions in both the $x$- and $y$-direction on \rosat PSPC images;
out to $8^\prime$ we accumulate spectra from four  
annular regions each $2^\prime$ wide; beyond this radius we accumulate  
spectra from annuli $4^\prime$ wide.  The energy dependent
PSF of the MECS and the energy-dependent telescope  
vignetting for on-axis observations have been taken into account in  
our extended sources analysis by generating appropriate instrument  
response files (with the {\it effarea} program available within the SAXDAS  
package) to be used when fitting the accumulated spectra. We have  
computed the corrected effective area for the $8^\prime-12^\prime$  
annulus, which is covered by the circular region of the strongback, by  
considering the typical thickness of the strongback and its  
transmission as a function of the energy and position. All other  
regions of the detector covered by the strongback have been  
appropriately masked and the data rejected.  The background  
subtraction has been performed using spectra extracted from blank sky  
events files in the same regions of the detectors as the source.  
  
We fitted each spectrum with a single-temperature plasma in  
collisional equilibrium at the redshift of the cluster   
({\sc Mekal} model --Kaastra 1992, Liedhal et al. 1995--   
in XSPEC v.~10.0 --Arnaud 1996),   
absorbed by the nominal Galactic column density ({\it wabs}  
model; Dickey \& Lockman 1990).

\section{Deprojection of the spectral results}  
  
The physical quantities constrained from fits of spectra with  
counts collected from cluster regions projected on the sky  
need to be converted to their values in the spherical shells   
that constitute the assumed spherical geometry of the X-ray emitting  
plasma. Fitting a thermal model to a projected spectrum  
provides, for each annulus,   
(i) an estimate for the Emission Integral, $EI = \int n_{\rm e}  
n_{\rm p} dV = 0.82 \int n_{\rm e}^2 dV$,   
through the normalization $K$ of the model, $K = \frac{10^{-14}}  
{4 \pi d_{\rm ang}^2 (1+z)^2} EI$ (see {\sc Mekal} model in XSPEC; 
we assume $n_{\rm p} = 0.82 n_{\rm e}$ in the ionized intra-cluster plasma);   
(ii) a direct measurement of the emission-weighted gas temperature,  
$T_{\rm ring}$ (note that the observed temperature is properly a
photon-weighted temperature, but the difference from our assumed
definition is completely negligible), 
metal abundance, $Z_{\rm ring}$, and   
luminosity, $L_{\rm ring}$.  
The purpose of the deprojection is, for example, to recover  
the value of the gas temperature in shells,   
$T_{\rm shell} \equiv T_i$, that is defined as  
\begin{equation}  
T_{\rm ring} \equiv T_j = \frac{ \sum_{i, {\rm outer \ shell}}^{i=j}  
T_i w_{ij} }{ \sum_{i, {\rm outer \ shell}}^{i=j} w_{ij} }  
\label{eq:tem}  
\end{equation}  
where $w_{ij} = L_i \times Vol(i,j)/Vol(i) = \epsilon_i Vol(i,j)$   
provides the luminosity  
for a given shell $i$ with volume $Vol(i)$ weighted by the part   
of this volume projected on the ring $j$, $Vol(i,j)$.  
Using this notation, it is simple to note that   
$L_{\rm ring} \equiv L_j = \sum_{i, {\rm outer \ shell}}^{i=j}   
\epsilon_i Vol(i,j) = \sum_{i, {\rm outer \ shell}}^{i=j} w_{ij}$.   
  
From Kriss, Cioffi \& Canizares (1983; see also McLaughlin 1999 
and, particularly relevant to X-ray analysis, Buote 2000),  
the volume shell observed through each ring adopted in the spectral  
analysis can be evaluated and a matrix, ${\bf Vol}$, can be built  
with components equal to the parts the volume of the shells (rows $i$)  
seen at each ring (or annuli; column $j$).  
  
The deprojected physical quantities can be then obtained through   
the following matrix products (shown by the symbol $\#$):  
\begin{equation}  
\begin{array}{l}  
n_{\rm e} = \left[ ({\bf Vol}^T)^{-1} \# (EI/0.82) \right]^{1/2} \\  
\epsilon = ({\bf Vol}^T)^{-1} \# L_{\rm ring} \\  
\epsilon T_{\rm shell} = ({\bf Vol}^T)^{-1} \# (L_{\rm ring} T_{\rm ring}) \\  
\epsilon Z_{\rm shell} = ({\bf Vol}^T)^{-1} \# (L_{\rm ring} Z_{\rm ring}),   
\end{array}  
\end{equation}  
where $({\bf Vol}^T)^{-1}$ indicates that the matrix is firstly  
transposed and then inverted.  
The emission due to the shells projected along the line of sight   
but with the corresponding annuli outside the field-of-view is taken   
into account with an edge correction factor estimated assuming a  
power law distribution of the emission proportional to $r^{-4}$  
(cfr. equation A8 in the appendix of McLaughlin 1999).  
Finally, we have to assign a single radius, $r_{\rm ave}$,
to each shell. Formally, for each shell delimited from the 
radii $r_i$ and $r_{i+1}$, this radius should be the one 
that solves the equation $n_{\rm gas}^2(r_{\rm ave}) = 3/(r_{i+1}^3-r_i^3)
\int_{r_i}^{r_{i+1}} n_{\rm gas}^2(r) r^2 dr$.
Considering that (i) $n_{\rm gas}^2(r) \propto r^{-\alpha}$ with 
$\alpha$ that has generally a value enclosed between 3 and 6 and can also vary 
between these values as function of radius in the same cluster, 
and (ii) complicated iterative procedure
and interpolation are required (see, e.g., discussion
about equation A9 in the appendix of McLaughlin 1999), we have checked
that the assumption of $r_{\rm ave} =(r_{i+1}+r_i)/2$ is consistent with
analytic solution of the equation above for an acceptable $\alpha$ 
within 2 per cent.

We have applied this technique to the single-phase results of the   
spectral analysis presented in DGM02.  
In rings where part of the flux was masked for the presence of  
point-sources, we correct the normalization $K$ by the relative  
amount of area not considered implicitly assuming  spherical  
symmetry.   
  
\begin{table*}  
\begin{center}   
\caption{Sample of 22 galaxy clusters considered in this study  
(see DGM02 for details on the reduction and spectral analysis  
of the \sax observations). The `CF' column indicates if a cluster   
is or not a cooling-flow system according to the amount of deposition  
rate quoted in Peres et al. (1998). $R_{\rm out}$ is the value   
of the radius at the outer end of the last radial bin considered.  
The best-fit parameters $r_s$ (in kpc) and $c$ for a King and a NFW 
dark matter density profile are quoted with the respective errors
(in parentheses). Note that A1367 and A3376 do not converge 
in the $r_{\rm s}-c$ plane.
} \label{tab1}  
\input{m-t_tab1_bestfit}  
\end{center}  
\end{table*}  
  
For each cluster, we finally have the following outputs:  
gas bolometric luminosity in each shell, $L \equiv L_{\rm shell}$;  
gas temperature in each shell, $T \equiv T_{\rm shell}$;  
electron density in each shell, $n_{\rm e}$, and, integrating  
it over the volume, the gas mass, $M_{\rm gas}$.  
An error is assigned to each quantity given the distribution  
of the values after 100 Monte-Carlo simulations obtained  
from scattering the original projected input with respect to their  
Gaussian error.  
  
Following DGM02, we divide our sample into  
two groups of objects with (CF; 12 clusters) and without  
(NCF; 10 clusters) a cooling flow in the central region   
(see reviews in Sarazin 1988, Fabian 1994),  
according to the mass deposition rate quoted in Peres et al.  
(1998; NCF systems have a deposition rate consistent with  
zero; cf. Table~\ref{tab1}).  
It is worth noticing, however, that this classification  
is meant to distinguish between relaxed (CF) and not-completely-  
relaxed (NCF) systems. In fact, also considering  
recent results from \chandra (David et al. 2001,  
Ettori et al. 2002, Sanders \& Fabian 2002, Johnstone et al. 2002) 
and  \xmm\ (Tamura et al. 2001, Molendi \& Pizzolato 2001,  
Matsushita et al. 2002)  
analyses of nearby clusters that show a lack of  
spectroscopic evidence of multi-phase gas,  
we parameterize all the X-ray emission with a single temperature  
model.  
  
\begin{figure}
\begin{center}
\epsfig{figure=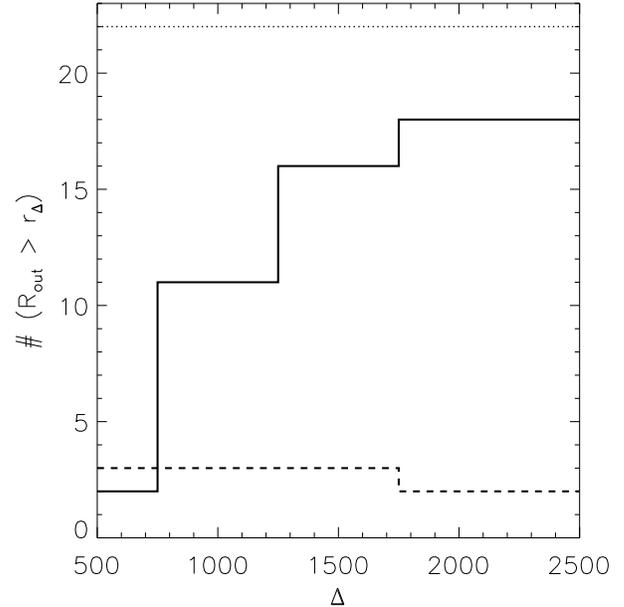,width=0.5\textwidth}
\caption{Histograms of the number of objects considered at each overdensity
$\Delta$ in our analysis. The {\it solid line} shows the
number of clusters where the outer radius, $R_{\rm out}$,
to which the quantities in exam are observed is larger than (or equal to)
$r_{\Delta}$ at the 95 per cent level of confidence 
[i.e. $R_{\rm out} \geq (r_{\Delta} - 1.96\sigma)$, where $\sigma$
is the error quoted in Table~\ref{tab:mass}]
as function of the overdensity $\Delta$.
The {\it dotted line} indicates the total number of clusters in our sample.
The {\it dashed line} shows the number of objects that do not satisfy the
selection criterion in mass (i.e. $M_{\rm tot} > 0$ at any given $r_{\Delta}$
at the 95 per cent level of confidence; see Sect.~5).
The number of objects considered at each $\Delta$ is given, therefore,
from ({\it dotted line} -- {\it dashed line}).
} \label{rout_rd} \end{center} \end{figure}

\begin{table*}
\begin{center}
\caption{Results from the deprojection analysis. 
All the quantities are estimated
within $R_{\Delta}$ (apart from $T$, 
which is estimated {\it at} $R_{\Delta}$), 
where the given overdensity $\Delta$
is obtained assuming either a King or NFW functional form
for the total mass profile ($1 \sigma$ error in parentheses).
} \label{tab:mass}
\input{m-t_tab_mass}
\end{center}
\end{table*}

\section{X-ray estimate of the gravitational mass profiles}  
  
To estimate the total gravitating mass, $M_{\rm tot}$,  
we make direct use of the deprojected gas temperature and electron   
density values estimated from the spectral best-fit with a   
single phase model. For each cluster, we select 
the mass model that reproduces better the deprojected gas temperature
profile inverting the equation of the hydrostatic equilibrium
between the dark matter potential and the intracluster plasma:
\begin{eqnarray}
-G \mu m_{\rm p} \frac{n_{\rm e} M_{\rm tot, model}(<r)}{r^2} =
\frac{d\left(n_{\rm e} \times kT\right)}{dr}  
\label{eq:mtot}
\end{eqnarray}
where $\mu$=0.6 is the mean molecular weight in a.m.u., $G$ is the
gravitational constant, $m_{\rm p}$ is the proton mass,
and using the deprojected electron density, $n_{\rm e}$.
As mass models, we consider two functional forms obtained from
the integration of the following dark matter density profile:
(i) the King approximation to the isothermal sphere (King 1962,
Binney \& Tremaine 1987), with a flat core in the inner part and
a $r^{-3}$ dependence at $r \rightarrow \infty$;
(ii) the function discussed in Navarro, Frenk \& White (1997, hereafter NFW), 
with a $r^{-1}$ and a $\sim r^{-2.4}$  dependence in the inner and outer 
parts, respectively:
\begin{eqnarray}
\lefteqn{ M_{\rm tot, model}(<r) = 4 \pi \ r_{\rm s}^3 \ \rho_{\rm s} \ f(x), }
\nonumber \\
\lefteqn{ \rho_{\rm s} =  \rho_{\rm c} \frac{200}{3} \frac{c^3}
{\ln (1+c) -c/(1+c)} },  \\
f(x) & = &  \left\{ \begin{array}{l}
\ln (x+\sqrt{1+x^2}) - \frac{x}{\sqrt{1+x^2}} \; {\rm (King)} \nonumber \\ 
\ln (1+x) - \frac{x}{1+x} \ \hspace*{1cm} {\rm (NFW)} 
\end{array}
\right.
\label{eq:mass_nfw}
\end{eqnarray}
where $x = r/r_{\rm s}$, $\rho_c$ is the critical density and
the relation $r_{\Delta=200} = c \times r_{\rm s}$ holds for the 
NFW profile.

Both of these mass models have two free parameters,
the core (King) or scale (NFW) $r_{\rm s}$ and the normalization, 
that we quote through the concentration parameter $c$
(note that we do this also for a King profile for convenience).
For each mass model, we obtain the best-fit values of the two parameters
minimizing the $\chi^2$ of the comparison between the deprojected 
temperature profile and the one obtained from eqn.~\ref{eq:mtot}
in two successive steps.
First, a minimum in a $\chi^2$ distribution is searched varying these  
parameters within the following ranges: 10 kpc $< r_{\rm s} <$  
max(2000 kpc, $R_{\rm out}$), 1 $< c <$ 15.  
This search provides the best-fit values $r_{\rm s}', c'$.  
A second fit is, then, performed in the restricted ranges:  
[min($r_{\rm s}'$-300 kpc, 10 kpc), max($r_{\rm s}'$+300 kpc, $R_{\rm out}$)],  
[min($c'$-3.0, 0.5), $c'$+3.0].  
The results of this refined fit on the scale radius $r_{\rm s}$   
and the concentration parameter $c$  
are presented in Table~\ref{tab1}.  
Hereafter, $M_{\rm tot}(<r)$ is defined for each cluster
according to the minimum $\chi^2$ provided from the 
two mass models considered, $M_{\rm tot, King}(<r)$ 
and $M_{\rm tot, NFW}(<r)$. 
The error related to the mass estimate is obtained 
from half the difference between the maximum and the minimum 
value calculated at each radius for the set of parameters 
acceptable at 1 $\sigma$.

From our final sample, we exclude A1367 and A3376 because
we do not obtain any $\chi^2$ solution for them.
Out of the remaining 20 objects (12 CF and 8 NCF systems), 
ten (6 of which are CF clusters) are fitted better with a 
King profile.

  
  
We investigate the relations among different physical quantities  
considering their values at a given overdensity, $\Delta$.  
This is defined with respect to the critical density,   
$\rho_{\rm c, z} = (3 H_z^2) / (8 \pi G)$, and   
within a cluster described as a sphere with radius $r_{\Delta}$:  
\begin{equation}  
\Delta = \frac{3 M_{\rm tot}(<r_{\Delta})}{4 \pi \rho_{{\rm c}, z}  
r_{\Delta}^3},  
\label{eq:delta}  
\end{equation}  
with the Hubble constant at redshift $z$ equal to  
\begin{equation}  
H_z = H_0 \sqrt{\Omega_{\rm m} (1+z)^3 +1 -\Omega_{\rm m} }  
\label{eq:hz}  
\end{equation}  
(for $\Omega_{\rm m}+\Omega_{\Lambda}=1$; i.e.,  
$H_0 \times (1+z)^{3/2}$ for an Einstein--de Sitter universe).  
  
The following analysis has been performed at different overdensities.  
To handle the observed profiles at any radius, we interpolate linearly
all the quantities on scales of 1 kpc.
In Fig.~\ref{rout_rd}, we show the number of clusters   
for which the region enclosing a given overdensity  
$\Delta$ is directly accessible to our X-ray observations.  
From this figure, we conclude that, at $\Delta=$2500,
18 galaxy clusters have a detectable X-ray emission and
two (A426 and A3526) need an extrapolation of the physical quantities
($R_{\Delta}/R_{\rm out} =$ 1.19 and 1.34 for A426 and A3526,
respectively). 
As reference value for our results (cf. Table~\ref{tab:mass}),   
we consider also $\Delta=1000$, where 11 objects
are observable, eight (A85, A119, A426, A496, A1656, A2029,
A3526, A3571) need an extrapolation in radius by about 40 per cent 
(5, 29, 87, 15, 39, 11, 95 and 20 per cent, respectively)
and one (A3627) did not satisfy our selection criterion 
in mass ($\sigma_{M} / M=$0.62 $>$ 0.51, see Sect.~5).
Furthermore, to compare our results with previous work,   
we estimate the quantities examined, i.e. gas density, 
temperature and luminosity, at lower overdensity.  
For those clusters without observed values at these $r_{\Delta}$,  
we extrapolate the interesting quantities using a least squares 
error-weighted fit with a first-order polynomial performed with the 
{\tt svdfit} function (Press et al. 1992, sect.~15.4)
on the logarithmic values of the variables observed 
in the outer region where $r > 0.7 \times R_{\rm out}$.
The mean relative error measured in the observed region is 
propagated to the extrapolated values.
  
For a given overdensity $\Delta$, we quote in Table~\ref{tab:mass}  
the values of $M_{\rm tot}(<r_{\Delta})$, $r_{\Delta}$,   
several estimates of the gas temperature (see Sect.~5),   
$L_{\rm bol} (<r_{\Delta})$ and $M_{\rm gas} (<r_{\Delta})$.

\subsection{X-ray mass: comparison with $\beta \gamma$--model}  
  
In this section, we compare the estimates of the dark matter profile we have  
obtained in the previous subsection   
with results derived from modelling the gas density profile   
and applying (i) the hydrostatic equilibrium and   
(ii) a polytropic shape of the temperature profile.
The latter procedure is generally applied in the X-ray analysis  
of galaxy clusters and makes use of the $\beta-$model,  
$\rho_{\rm gas} \propto (1+x^2)^{-1.5 \beta}$ 
(Cavaliere \& Fusco-Femiano 1976),  
to reproduce the observed surface brightness profile   
(an analytic expression can be obtained if $T_{\rm gas}$ is assumed constant;  
for a generalization to the polytropic case see Ettori 2000)  
and build a temperature profile as function of the polytropic index  
$\gamma$ like $T_{\rm gas} \propto (1+x^2)^{-1.5 \beta (\gamma-1)}$.    
  
In the $\beta \gamma$--model the total mass profile is readily derived  
from   
eqn.~\ref{eq:mtot}  
\begin{eqnarray}  
\frac{M_{\rm tot, \beta \gamma} (< r)}{10^{14} h^{-1}_{50} M_{\odot} }  
 & = & 1.11 \ \frac{0.6}{\mu} \ \beta \ \gamma \ T(r) \ r_{\rm c} \   
  \frac{x^3}{\left( 1+x^2 \right)}  \nonumber \\  
 & = & 1.11 \ \frac{0.6}{\mu} \ \beta \ \gamma \ T_0   
\ r_{\rm c} \ \frac{x^3}{\left( 1+x^2 \right)^{1.5 \beta (\gamma-1)+1}}   
\label{eq:mbeta}  
\end{eqnarray}  
  
This formula has been applied to estimate the total mass profile  
in recent work that considered a measured temperature profile from  
\asca data (e.g. Markevitch et al. 1999 on A496 and A2199;   
Nevalainen, Markevitch \& Forman 2000,   
Finoguenov, Reiprich \& B\"ohringer 2001).   
  
\begin{figure*}  
\hbox{  
\epsfig{figure=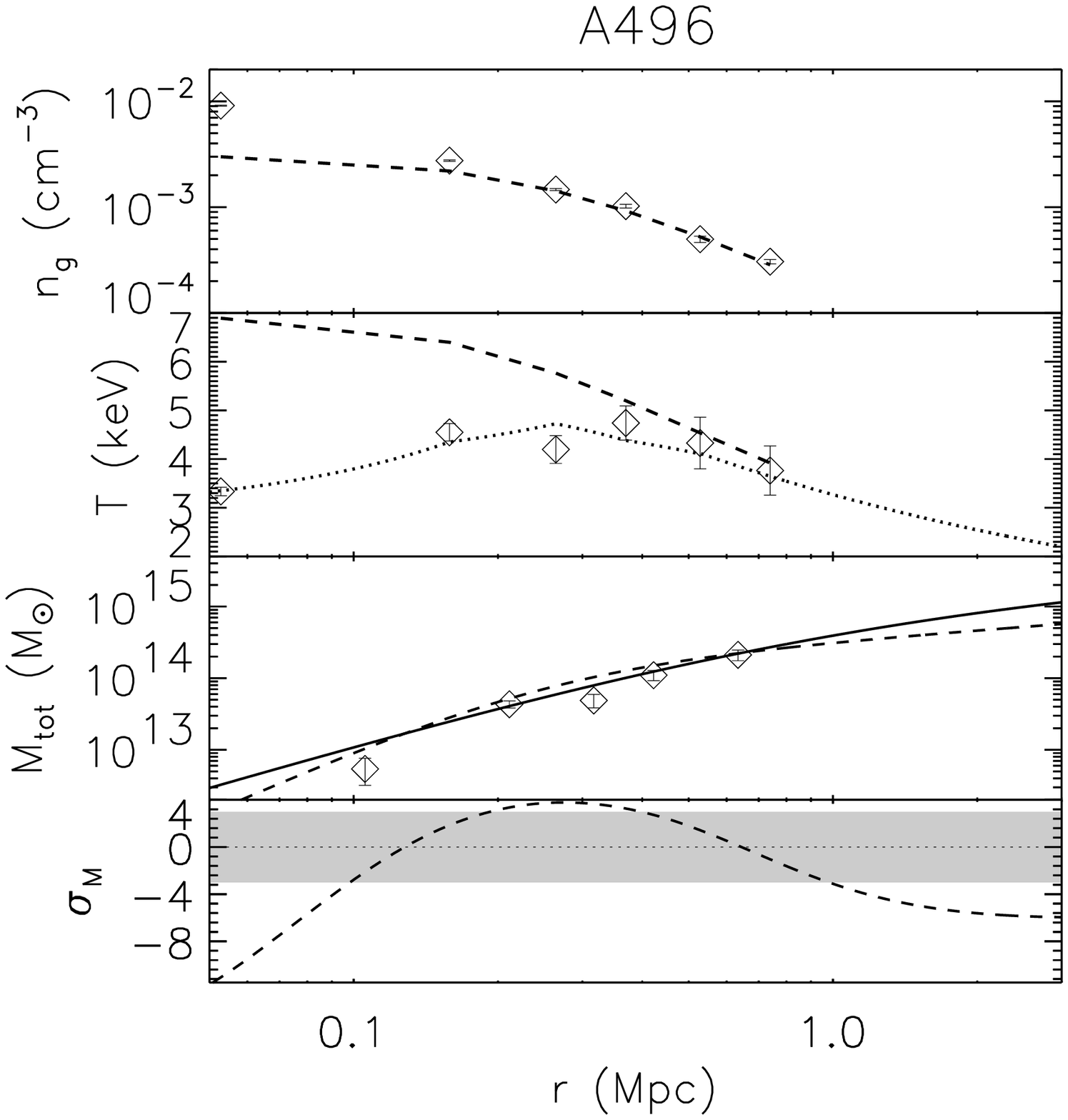,width=0.5\textwidth}  
\epsfig{figure=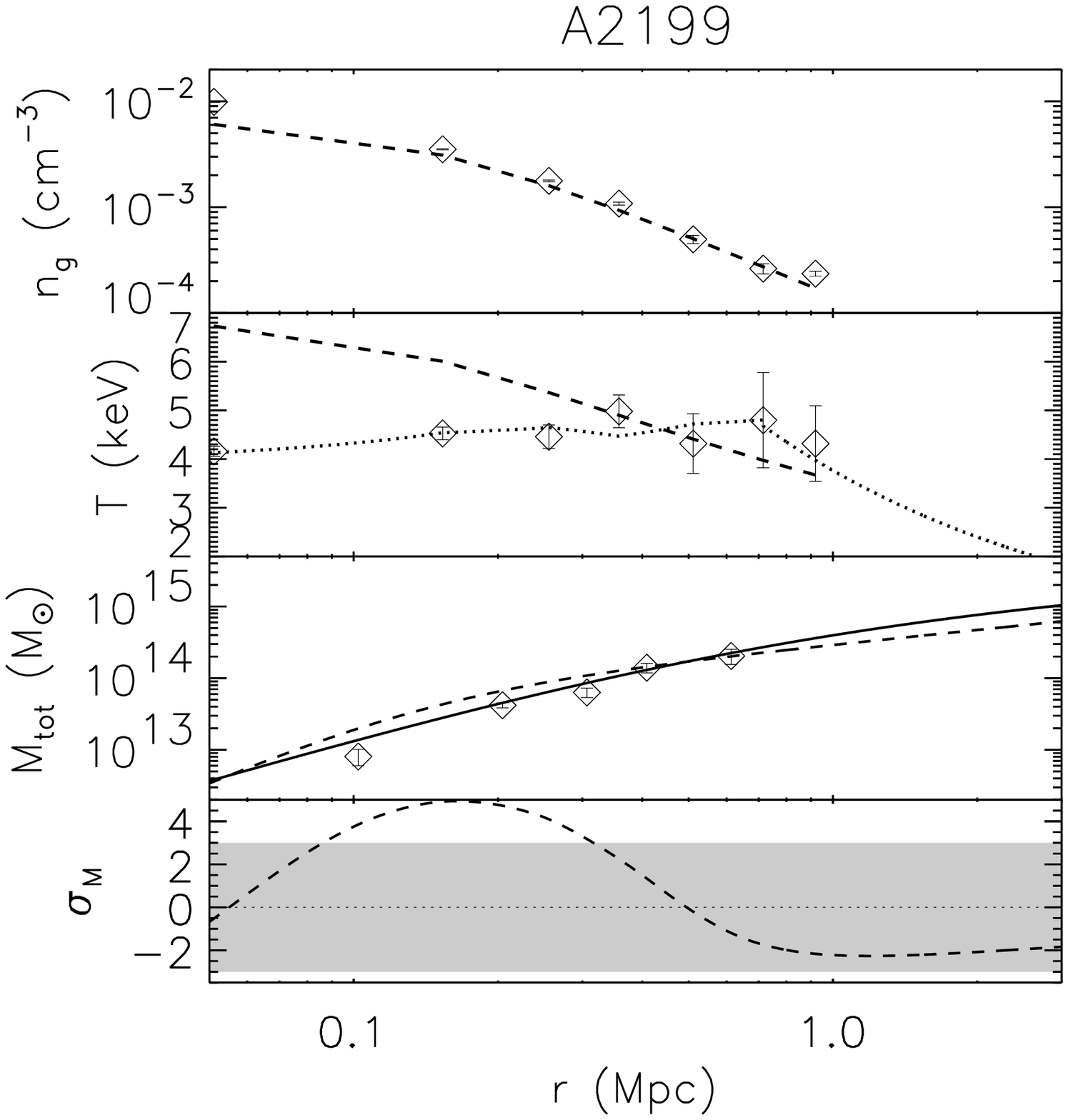,width=0.5\textwidth}  
} \caption{Comparison between deprojected (diamonds) 
observed values and best-fit $\beta \gamma$--model (dashed line)  
for gas density, temperature and total gravitating mass.  
The dotted line indicates the best-fit temperature profile
for a given mass model as described in Sect.~4.
} \label{model} \end{figure*}

In fig.~\ref{model}, we compare our mass profiles 
(from the King functional form, cf. Table~\ref{tab1}) 
for two CF clusters (A496 and A2199) with those derived
by Markevitch et al. (1999) using the $\beta \gamma$ model. 
The larger deviations ($|\sigma_{\rm M}| > 3 \sigma$) are localized   
in the region $\sim 100-500$ kpc (and below 100 kpc and above 1 Mpc
in A496)
and introduce a systematic error that could contribute to
the observed scatter in the distribution of the measurements.   
It is worth noticing that the polytropic temperature profile does
not reproduce in a satisfactory way the temperature profile of either A496
or A2199, as can be seen in fig.~\ref{model}.
DGM02 have shown that the temperature profiles of our \sax sample of
clusters are in general not in good agreement with polytropic 
temperature profile. This discrepancy should be taken into consideration 
when applying the $\beta \gamma$ model to derive mass measurements 
of galaxy clusters.

\subsection{X-ray mass: comparison with optical estimates}  
  
\begin{figure}  
\epsfig{figure=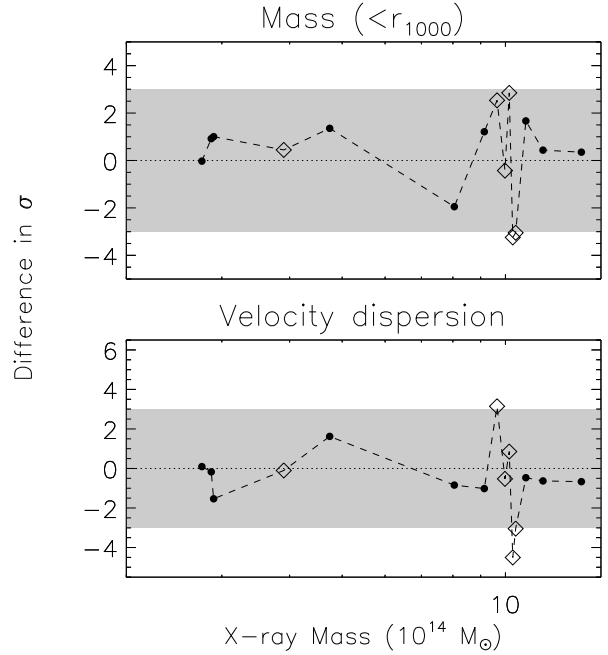,width=0.5\textwidth}  
\caption{Differences (in $\sigma$) between the interpolated  
optical values and the measured X-ray mass and velocity dispersion  
versus the measured X-ray mass (see text for details).   
{\it Filled circles} represent CF galaxy clusters, whereas  
{\it open squares} are NCF objects.  
} \label{m_opt} \end{figure}

Girardi et al. (1998) quote the optically-determined mass estimates  
for 15 out of 20 of our clusters (not available for A3526, A3627,
2A0335, PKS0745, TRIANG).   
In Fig.~\ref{m_opt}, we show a comparison between X-ray measurements   
at $\Delta=$1000 and the optical masses estimated at the same $r_{\Delta}$  
by using the Jeans equation ($M_{\rm iso} = 3 \beta_{\rm gal}  
\sigma_{\rm p}^2 r /G$, where $\beta_{\rm gal}$ is the exponent in the  
King-like galaxy density profile and $\sigma_{\rm p}$ is the projected   
velocity dispersion. This mass estimate is consistent with the corrected   
virial mass as discussed in Girardi et al. (1998, Section~5).  
Moreover, by making use of the relation $\sigma^2 = (G M_{200})/(2 r_{200})$  
between the velocity dispersion in the dark matter distribution and  
the total mass within $r_{200}$, we derive   
$\sigma_X = \sqrt{50} \ H_0 \ r_{200}   
= \sqrt{50} \ H_0 \ c \ r_{\rm s}$   
for a NFW potential and compare our estimates of $\sigma_X$  
from the best-fit values in Table~\ref{tab1} to the optically-determined  
velocity dispersion, $\sigma_{\rm p}$.    
   
Out of 16 clusters examined, we observe two systems (A119 and A754)
lying with a relative difference in mass larger than
3 $\sigma$. When the optically and X-ray determined
velocity dispersions are compared, three objects (A119, A754 and A2256)
show significant deviations (see Fig.~\ref{m_opt}).
These three clusters are NCF systems, are 
known to have irregular and asymmetric X-ray brightness
and, at least for A754 and A2256, are indeed involved 
in massive merging (e.g., A119: Ferretti et al. 1999;
A754: Henriksen \& Markevitch 1996;
A2256: Molendi, De Grandi \& Fusco-Femiano 2000) that may affect
both the optical determinations of the velocity dispersion  
and the validity of the hydrostatic assumption made  
in the process of the estimation of the X-ray mass.  
  
In general, we measure a median deviation of about  
1.2 and 0.8 $\sigma$ in mass and velocity dispersion  
measurements, respectively. 
Moreover, there is evidence that larger deviations
are present in the subsample of NCF, not-relaxed systems
($2.7$ and $2.0 \sigma$ deviation in mass and   
velocity dispersion, respectively, for NCF;  
$1.0$ and $0.7 \sigma$ for CF).  

\section{Relations among the observed quantities}
  
Our refined sample of 20 nearby (0.010 $< z <$ 0.103; 
median redshift of 0.050) clusters of galaxies spans 
a factor of nore than three in mass-weighted temperature   
(3.1 keV $< T_{\rm X} <$ 9.9 keV; median value: 6.9 keV) and   
two orders of magnitude in luminosity ($1.7 \times 10^{44}$ erg s$^{-1}$   
$< L_{\rm X} < 6.1 \times 10^{45}$ erg s$^{-1}$; median value:  
$1.5 \times 10^{45}$ erg s$^{-1}$).  
In the following analysis, we consider only the clusters with a total
mass at a given radius larger than zero at the 95 per cent level
of confidence, i.e.
we select just the objects with $\sigma_{M} / M < (1/1.96) = 0.51$. 
  
In the present work, we investigate the correlations between  
the observed physical quantities in order  
to assess the robustness of the self-similar scaling relations
for clusters of galaxies (e.g. Kaiser 1986).  
These relations are the product of simple assumptions on the formation
and evolution of galaxy clusters. As non-linear structures,
they are assumed to form by homogeneous spherical collapse of 
gravitational instabilities of dark matter on which gas infalls are
heated up by shocking processes. If no dissipation is considered,
adiabatic X-ray emitting plasma can be considered to share
the same potential well with dark matter with a spatial distribution 
that can be different from the dark matter's one but has to be the same
at any earlier epoch (e.g. Bryan \& Norman 1998, Arnaud \& Evrard 1999).  
However, deviations are expected from self-similarity under  
the effects of, for example, the dynamical history of clusters   
as three-dimensional aggregation of clumps (Jing \& Suto 2000,  
Thomas et al. 2001) and any additional physics acting on the   
intracluster gas over the simplistic infall in the potential well  
(e.g. Evrard \& Henry 1991; David, Forman \& Jones 1991;  
Bryan \& Norman 1998; Bialek, Evrard \& Mohr 2001; Borgani et
al. 2002 and references therein).  
The latter case is particularly relevant to cool systems where  
the extra energetic amount required from their observed properties  
is comparable to their thermal energy  
(e.g. Ponman et al. 1996; Ponman, Cannon \& Navarro 1999;   
Tozzi \& Norman 2001).  
Our sample encloses only clusters with temperature larger  
than about 3 keV and, thus, is expected not to be affected   
in a significant way by any increase of the gas entropy   
occurring during the cluster formation history.  
Therefore, we are able to investigate the  
galaxy cluster scaling laws excluding systematics  
related to the energetic budget. 

\begin{table}  
\begin{center}   
\caption{Spearman's $\rho$ rank correlation results on a set of  
physical quantities. A small value in probability indicates   
significant correlation. $n \sigma$ indicates the number   
of standard deviations by which the dependence in exam  
deviates from the null-hypothesis of uncorrelated data sets.  
The gas-mass-weighted temperature, $T_{\rm mw}$, is here used.
} \label{tab2}  
\input{m-t_tab2.tex}  
\end{center}  
\end{table}  
  
\begin{figure*}  
\hbox{  
\epsfig{figure=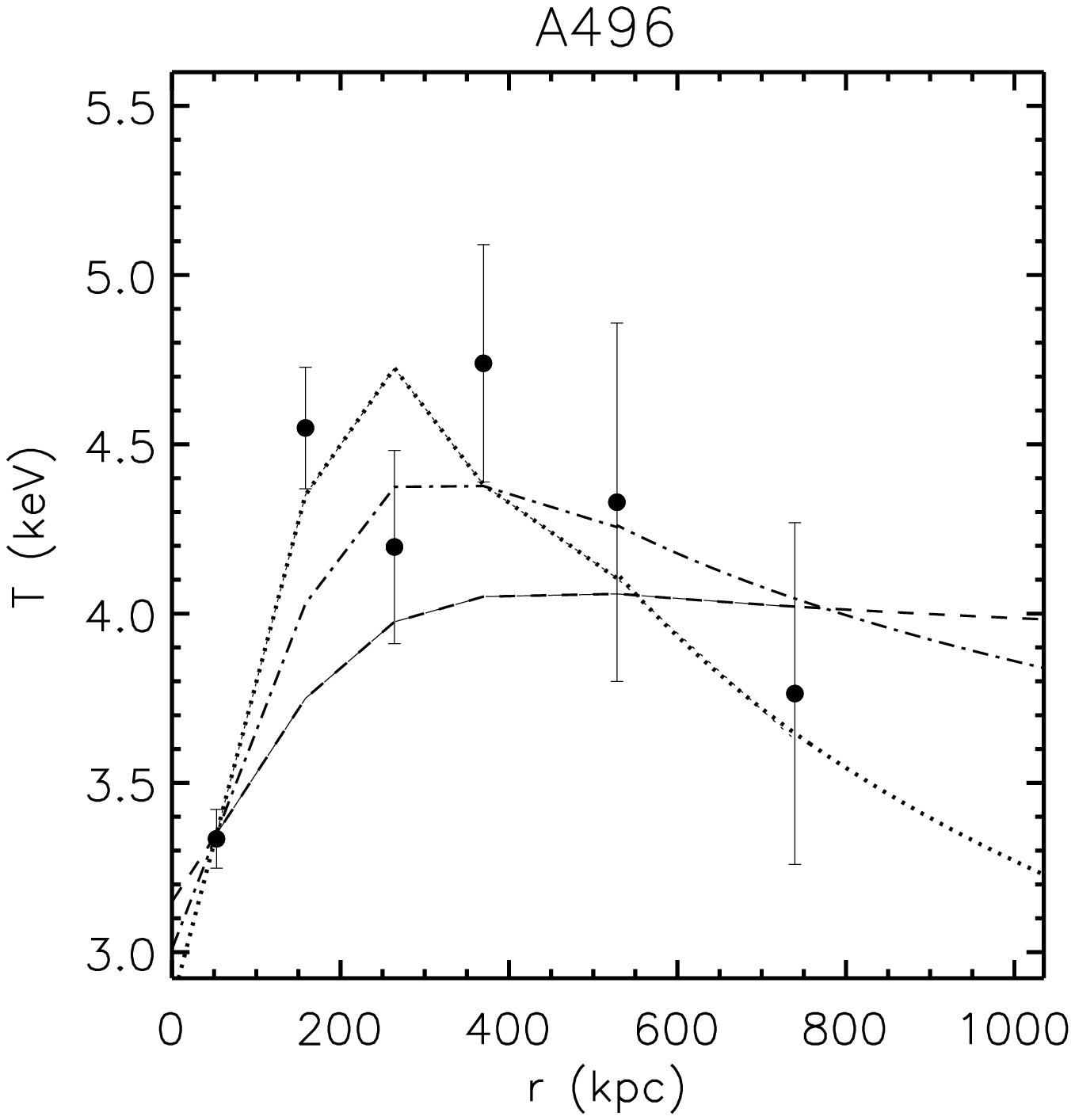,width=0.5\textwidth,angle=0}  
\epsfig{figure=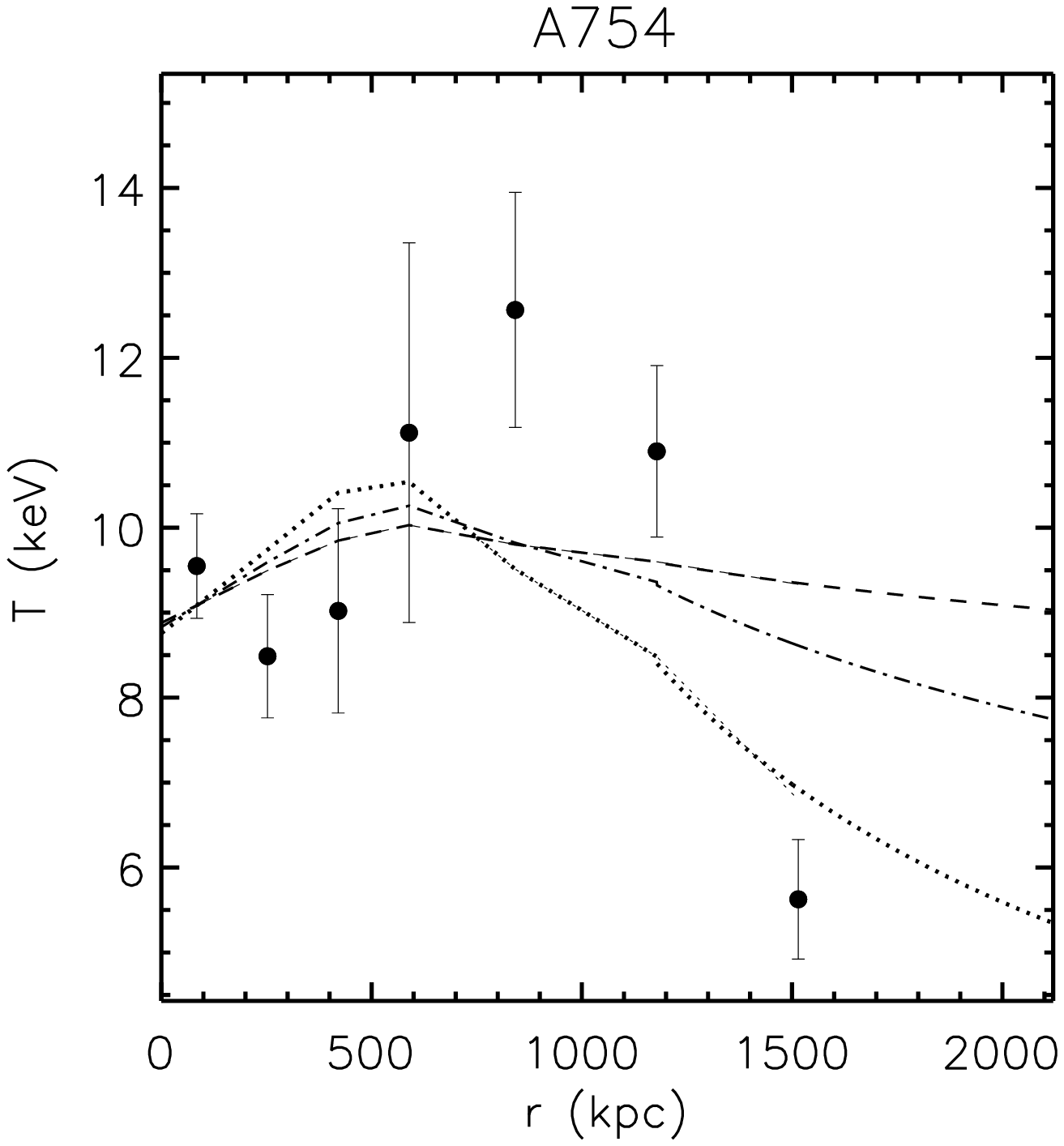,width=0.5\textwidth}  }  
\caption{  
The temperature defined in eqn.~8, 9 and 10 are here plotted  
({\it dotted}, {\it dashed} and {\it dot-dashed}, respectively)   
with the de-projected data points.
} \label{fig:tt} \end{figure*}

In our analysis, we adopt three different definitions  
for the plasma temperature at a given overdensity $\Delta$:  
 
\begin{enumerate}  
\item the gas temperature in the shell {\it at} $r_{\Delta}$,  
\begin{equation}  
T(r_{\Delta}) = T_i(r_{\Delta}),  
\end{equation}  
  
\item the emission-weighted gas temperature {\it within}  
$r_{\Delta}$,   
\begin{equation}  
T_{\rm ew}(<r_{\Delta}) = \frac{ \sum_i^{0<r_i<r_{\Delta}}  
 L_i \ T_i }{ \sum_i^{0<r_i<r_{\Delta}} L_i } =  
 \frac{ \sum_i^{0<r_i<r_{\Delta}} L_i \ T_i }{L (<r_{\Delta})},  
\end{equation}  
  
\item the mass-weighted gas temperature {\it within}   
$r_{\Delta}$,   
\begin{equation}  
T_{\rm mw}(<r_{\Delta}) = \frac{ \sum_i^{0<r_i<r_{\Delta}}  
 M_{\rm gas, i} \ T_i }{ \sum_i^{0<r_i<r_{\Delta}} M_{\rm gas, i} } =  
\frac{ \sum_i^{0<r_i<r_{\Delta}} M_{\rm gas, i} \ T_i }  
{M_{\rm gas}(<r_{\Delta})},  
\end{equation}  
\end{enumerate}  
where $i$ indicates the running cursor on shells and  
$j$ on rings. 
We use the values of the temperature in the volume shells
obtained from the best-fit procedure discussed in Sect.~4.
In Fig.~\ref{fig:tt}, we show the temperature profiles defined above  
for a typical CF (A496) and NCF (A754) galaxy cluster.  
  
To check for the presence of linear dependence   
between the logarithmic values of the physical quantities 
being studied,   
we have performed  both a non-parametric and a parametric  
analysis of our data.  
The two approaches can be considered as complementary:  
the non-parametric analysis has the advantage of not  
relaying on a specific model but does not treat errors; 
the parametric analysis, which assumes a power-law model,  
provides a treatment for errors.    
 
For the non-parametric analysis we have used  
the Spearman's $\rho$ rank correlation of two   
sample populations (Press et al. 1992, p.634) .   
We quote in Table~\ref{tab2} the set of pair quantities with   
the respective Spearman's $\rho$ and probability.  
Relations with a $\sim 3 \sigma$ deviation from the   
null hypothesis of uncorrelated datasets are present  
amongst the gravitating mass, the gas temperature,   
the gas luminosity and the gas mass. A weaker dependence  
appears between the gas mass fraction and the temperature.  
  
In our parametric analysis, we use the bisector modification
(i.e. the best-fit results bisect those obtained from minimization
in vertical and horizontal directions) of the linear regression
algorithm in Akritas \& Bershady (1996 and references therein,
hereafter BCES) that takes into account
both any intrinsic scatter
and errors on the two variables considered as symmetric.
The uncertainties on the best-fit results are obtained
from 10,000 bootstrap resamplings. 
  
\subsection{$M_{\rm tot} - T$ relation}  
 
In this and the following subsections we shall compare the  
the normalization and slope of the scaling relations  
obtained from our data to theoretical predictions
based on the simplistic assumption of an  
isothermal sphere for both the gas represented by its temperature   
and the collisionless dark matter particles (e.g. Kaiser 1986,  
Bryan \& Norman 1998). As usual we indicate with   
$\beta_T = (\mu m_{\rm p} \sigma^2)/(kT_{\rm gas})$, the ratio between  
the energy   
in the plasma and in the dark matter with velocity dispersion, $\sigma$.  
An isothermal distribution function is characterized by a proportional  
relation between the matter density and the velocity dispersion,   
$\rho(r) \propto \sigma^2/r^2$ (Binney \& Tremaine 1987).  
This implies a total mass within a radius $r$ of $(2 \beta_T \ kT)  
/(G \mu m_{\rm p}) \ r$ that can be compared to eqn.~\ref{eq:delta}  
to infer the relation between $M_{\rm tot}$ and $T_{\rm gas}$ at given  
overdensity $\Delta$:  
\begin{eqnarray}  
\lefteqn{ \frac{M_{\rm tot} (< r_{\Delta})}{10^{14} h^{-1}_{50} M_{\odot} }  
  = 0.38 \ \beta_T^{3/2} \ \left( \frac{50}{H_z}\right) \  
  \left( \frac{1000}{\Delta} \right)^{1/2} \  
  \left( \frac{T_{\rm gas}}{1 \mbox{keV}} \right)^{3/2}, } \nonumber \\ 
\lefteqn{ \log M_{14} = -0.42 \ +1.5 \log T \left(+\log 
  \left(\beta_T^{3/2} f_{\Omega} \right) \right), }  
\label{eq:m-t}  
\end{eqnarray}  
where $H_z$ is the Hubble constant at redshift $z$ given in   
eqn.~\ref{eq:hz}, $f_{\Omega} \equiv \left( \frac{50}{H_z}\right) \  
\left( \frac{1000}{\Delta} \right)^{1/2}$ and in the bottom row we have  
rewritten the relation in log-log form.  
The assumption on the dark matter profile only affects the value  
of the normalization. We consider the isothermal case here adopted  
as a reference.  
  
In the considerations above, we have adopted the assumption that we are  
observing clusters just after their virialization (cfr. Voit \& Donahue 1998  
for the implication on the $M-T$ relation of clusters that gradually form  
and stop evolving in a low density Universe).  
The $M-T$ relation makes reasonable assumptions that have been tested   
both in numerical simulations and in observations.  
Moreover, this is a direct result coming from the combination of the  
conservation of energy throughout nearly-spherical collapse of clusters  
with the virial theorem (Afshordi \& Cen 2002).  
  
\begin{table*}  
\begin{center}   
\caption{Results of the best-fit analysis.  
When the value of the slope is investigated,   
we apply the linear BCES bisector  
estimator to the logarithmic of the power law $Y = a X^b$,  
$\log Y = A +B \log X$ (i.e. $a = 10^A, b=B$; errors in parentheses).  
The temperature, $T$, is in unit of keV; the luminosity, $L$, in  
$10^{44} h_{50}^{-2}$ erg s$^{-1}$; the total mass, $M$, in   
$10^{14} h_{50}^{-1} M_{\odot}$; the gas mass, $M_{\rm gas}$, in  
$10^{13} h_{50}^{-5/2} M_{\odot}$; the radius at given overdensity,  
$R$, in 100 $h_{50}^{-1}$ kpc; the gas fraction, $f_{\rm gas}$,  
in $h_{50}^{-3/2}$.   
When the slope $B$ is fixed, we estimate the median of the  
distribution of $\log Y -B \log X$.   
The scatter on $Y$ is measured as $\left[ \sum_{j=1,N}   
\left(\log Y_j -A -B \log X_j \right)^2 /N \right]^{1/2}$.  
Note that the scatter along the X-axis can be estimated as   
$\sigma_{\log X} = \sigma_{\log Y} / B$. }   
\label{tab:fit}  
\input{m-t_tab_fit}  
\end{center}  
\end{table*}

\begin{figure*}
\hbox{
\epsfig{figure=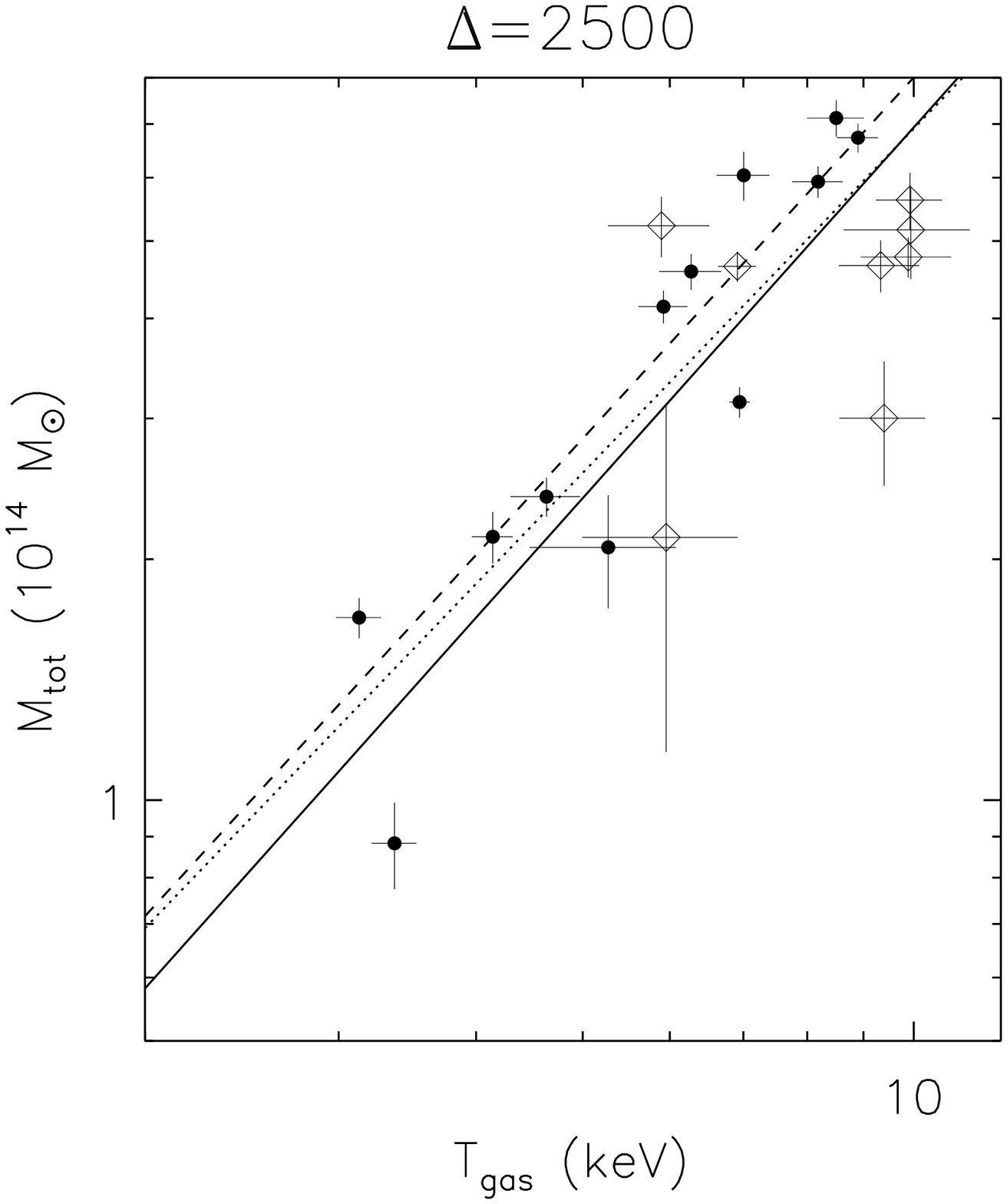,width=0.5\textwidth}  
\epsfig{figure=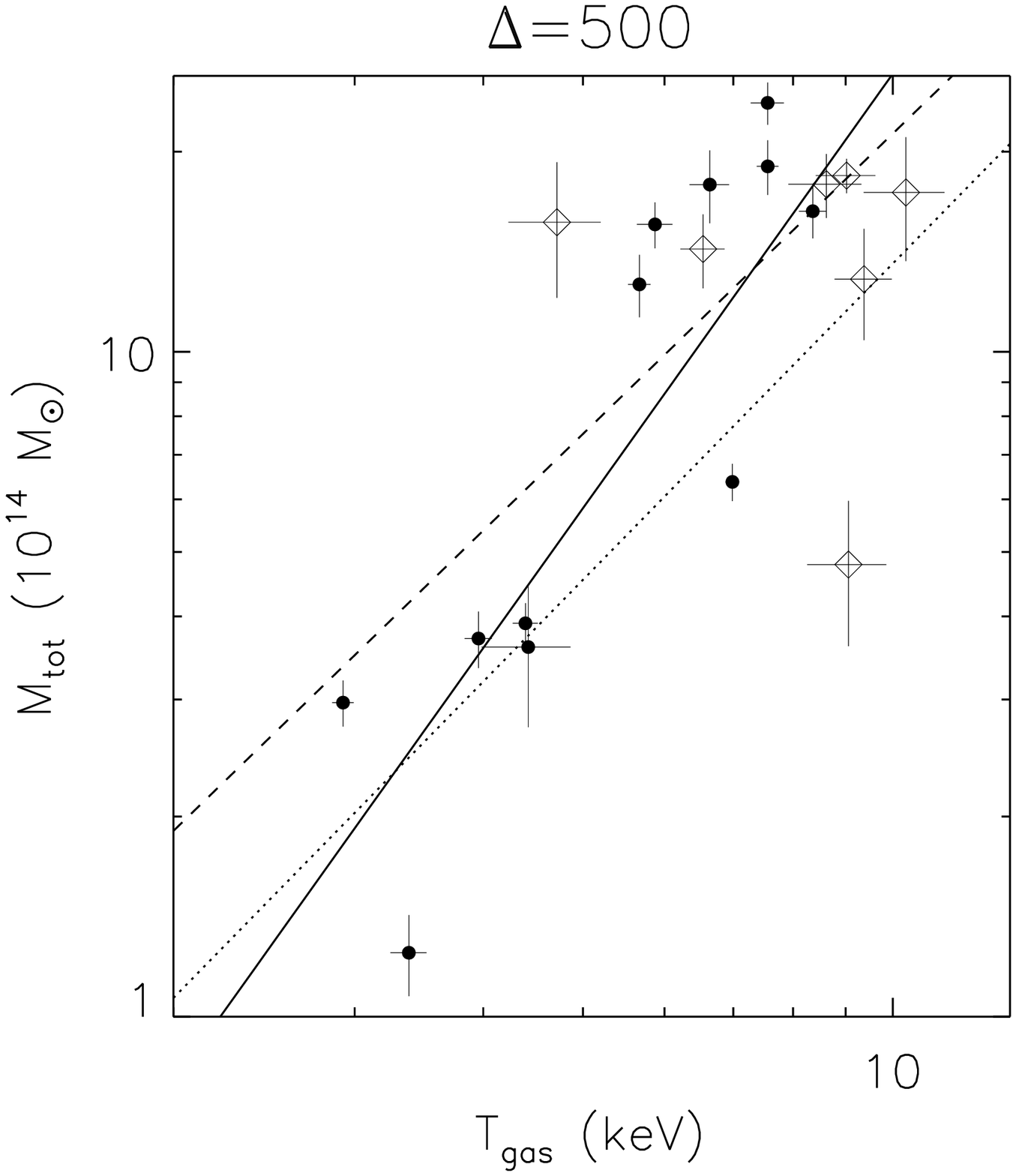,width=0.5\textwidth}  }
\caption{$M-T_{\rm mw}$ relation.
The {\it solid} and {\it dashed} lines represent our best-fit
results for the given overdensity, using the slope as free parameter
and fixing it to 1.5, respectively.
{\it Filled circles} represent CF galaxy clusters, whereas
{\it open squares} are NCF objects.
(Left) The best-fit slope is $1.54 (\pm 0.22)$.
The {\it dotted} line represents the best-fit result 
from Allen et al. (2001; $\Delta=2500$).
(Right) Best-fit of the $M-T_{\rm ew}$ relation at $\Delta=500$
(slope: $2.17 \pm 0.37$). The {\it dotted} line shows 
the best-fit from Finoguenov et al. (2001).
} \label{m_t} \end{figure*}

\begin{figure}  
\epsfig{figure=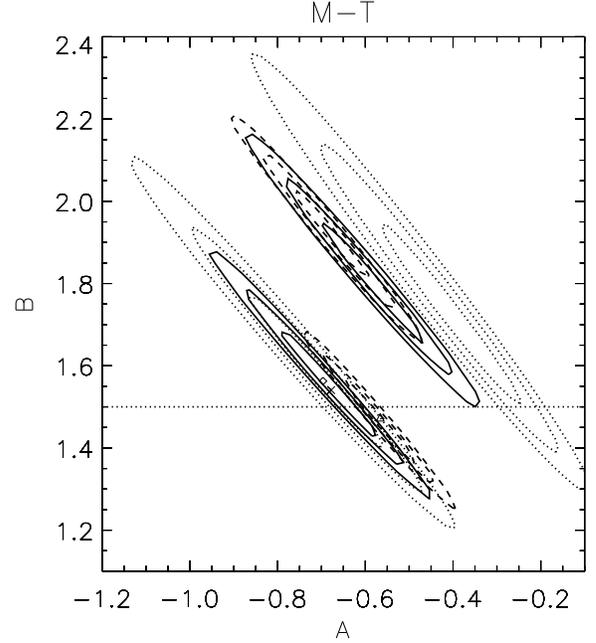,width=0.5\textwidth}    
\caption{Plot of 1, 2, 3 $\sigma$ contour from $\chi^2$ statistic 
for the two interesting parameters of the linear fit 
applied to $M-T_{\rm mw}$ (solid line),
$M-T_{\rm ew}$ (dashed line), $M-T(R)$ (dotted line) relations
at $\Delta=$2500 and 1000 (the latter ones have higher normalization $A$).
} \label{m_t_cont} \end{figure}

\begin{figure*}  
\hbox{
\epsfig{figure=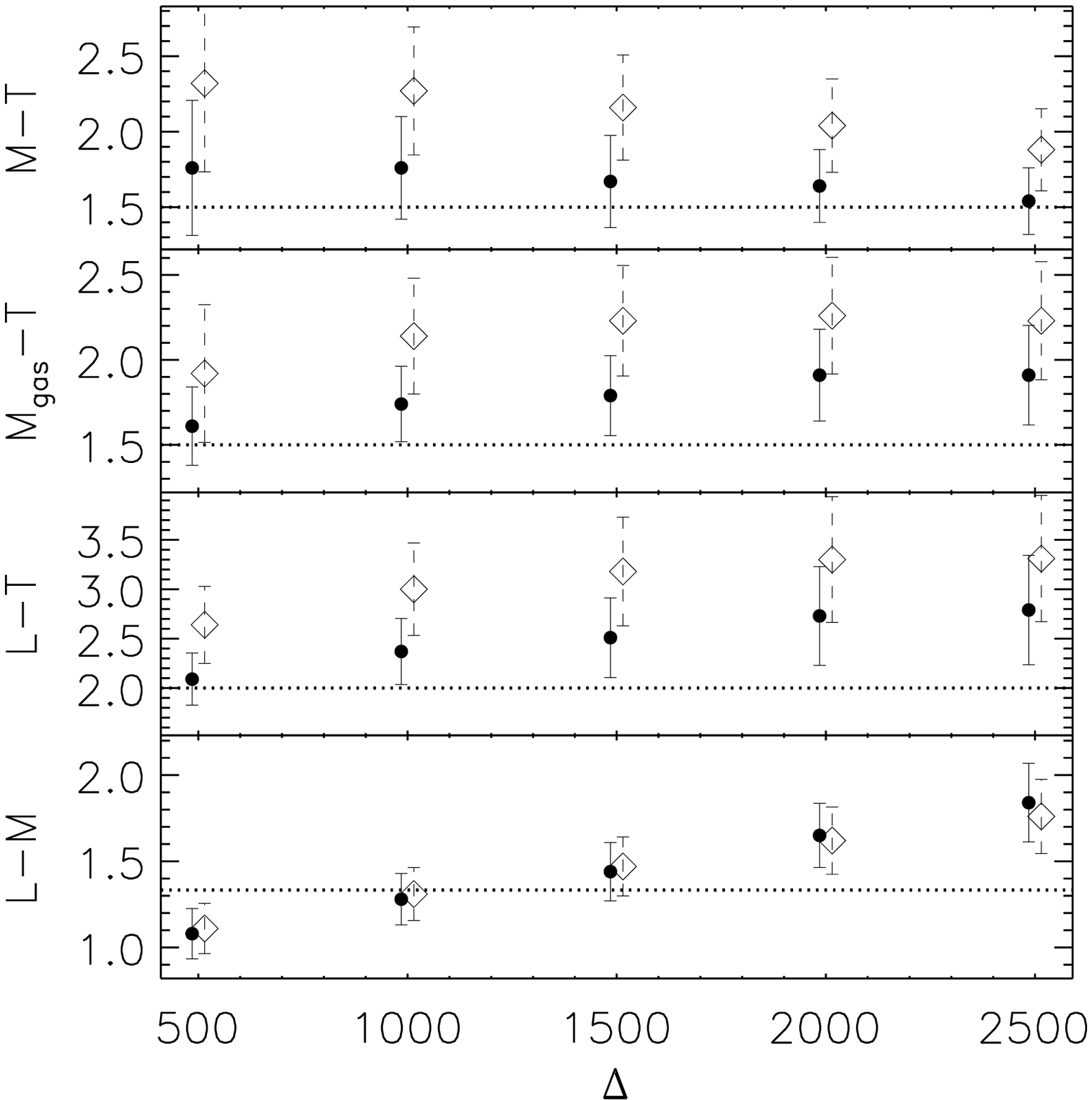,width=0.5\textwidth,angle=0}  
\epsfig{figure=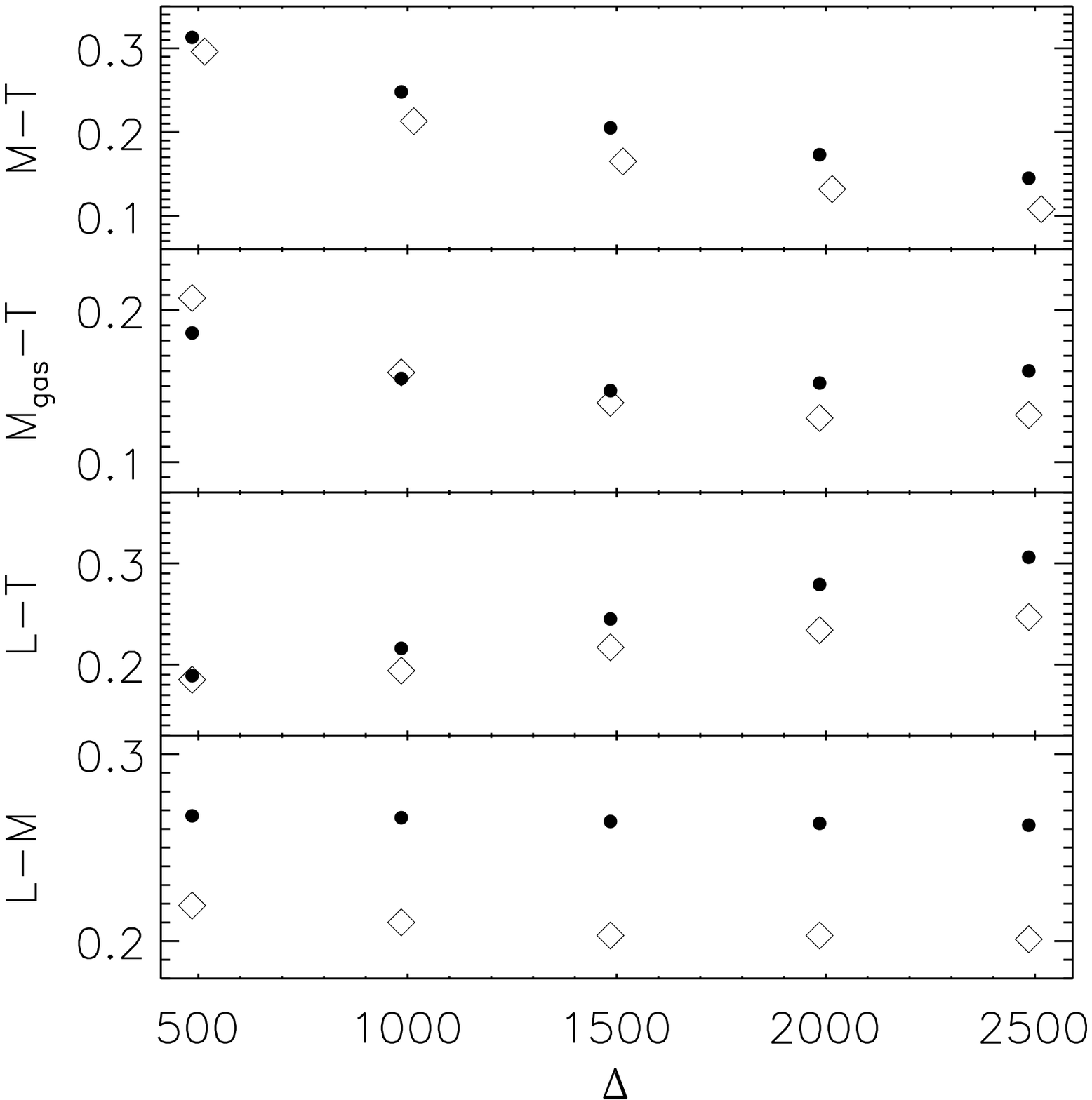,width=0.5\textwidth,angle=0}  
} \caption{(Left) Behaviour of the slope in the   
$M-T_{\rm mw}$, $M_{\rm gas}-T_{\rm mw}$, $L-T_{\rm mw}$ and $L-M$  
relations as function of the considered overdensity.  
The {\it diamonds} + dashed error bars show the results for CF systems only.
The dotted lines indicate the values predicted from the scaling  
laws. (Right) Values of the scatter of the same relations.
{\it Full dots} represent the entire sample, whereas the {\it diamonds}
indicate the results for CF objects. 
} \label{rel_dens} \end{figure*}

In Table~\ref{tab:fit} and Fig.~\ref{m_t},   
we show the results of the fitting analysis.  
A segregation is noticeable between CF and NCF objects.  
When we fit the twelve CF clusters, we measure   
$M_{14} = 0.12 (\pm 0.06) \times T_{\rm mw}^{1.88 (\pm 0.27)}$.  
When only the 8 NCF systems are considered,   
$M_{14} = 0.92 (\pm 2.01) \times T_{\rm mw}^{0.73 (\pm 1.00)}$.  
The slope does not show any significant change at the   
variation of the overdensity at which the quantities   
examined are considered (see Fig.~\ref{rel_dens}).  
  
We compare now these results on the normalization and slope of the $M-T$   
relation with the values obtained in previous work.  
  
In numerical simulations, fixing the slope to 3/2,   
the normalization $\beta_T$ in  eqn.~\ref{eq:m-t}  
ranges between 1.15 (model {\it CL2} in   
Navarro, Frenk \& White 1995) to  
1.24 (Evrard, Metzler \& Navarro  1996) and  
$\sim 1.3$ (Bryan \& Norman 1998; cfr. their Table~2).  
We measure a normalization $\beta_T$ of $1.14 (\pm 0.34)$  
that is consistent with the results quoted for 
simulated clusters.
For example, we obtain a best-fit correlation of   
$M_{14} = 0.46 (\pm 0.21) \times T_{\rm mw}^{1.5}$ ($\Delta=1000$),  
that has a normalization lower by 13 per cent 
(and only $0.3 \sigma$ apart) than the value   
measured at the same overdensity in the gas-dynamic simulations   
of Evrard et al. (1996).   
   
From an observational point of view,   
Horner, Mushotzky \& Scharf (1999) claim   
that the $M-T$ relation is steeper than the traditional scaling,  
following a $\propto T^{1.8-2.0}$ law,   
when the mass is estimated according to the $\beta-$model.  
Slopes steeper than virial prediction are also observed   
in high-redshift clusters (Schindler 1999)   
and highly-luminous clusters (Ettori \& Fabian 1999) samples,  
where isothermality is assumed.  
Moreover, Neumann \& Arnaud (1999) from a $\beta-$model estimate of the  
gravitating mass obtain a $M-T$ relation consistent  
with the classical scaling relation.   
Nevalainen, Markevitch \& Forman (2000)   
from a sample of 6 clusters  
and 3 groups/galaxies with temperature profiles  
observed from \asca and \rosat, respectively, inferred at  
$\Delta=1000$ a slope of $1.79 (\pm 0.09)$ and a normalization  
significantly lower than the one observed in simulations, suggesting  
evidence of breaking of the self-similarity in the  
less massive systems due to heating processes.  
Finoguenov, Reiprich \& B\"ohringer (2001)  
studied two samples of clusters,  
one comprising a complete sample of 63 bright objects from  
the \rosat All Sky Survey with an assigned emission-weighted temperature  
collected from the literature and the other including 39 systems  
(22 of these with $T <$ 3.5 keV) with known temperature  
profiles that are used to infer the total mass in combination  
with a $\beta \gamma$--model (see Section~4.1 above).  
Correlating $M_{\rm tot} (<r_{500})$ with an emission-weighted temperature,  
these authors find a slope of $1.58 (\pm 0.07)$  
for the sample with a resolved temperature profile and excluding objects  
with $M_{500} < 5 \times 10^{13} M_{\odot}$ (the slope increases slightly  
to $1.78 \pm 0.09$ for the whole sample, consistent  
with the result from the flux limited sample).  
The normalization is more than 50 per cent lower than the value  
quoted in Evrard et al. (1996).  
Allen, Schmidt \& Fabian (2001), using spatially resolved X-ray  
spectroscopy with the \chandra observatory  
of five highly massive cooling-flow  
(and so, relaxed) galaxy clusters at intermediate  
redshifts, found consistency with the scaling law prediction and  
a normalization 40 per cent lower (but with a deviation   
significant only to $1.8 \sigma$ considering their quoted  
values) than what is observed in simulated clusters.  
  
It is worth noticing that  a tight correlation between  
the parameters $A$ and $B$ of the linear fit performed  
on the logarithmic values appears from the   
contour plot of the probability distribution   
obtained applying the $\chi^2$ statistic   
(see Fig.~\ref{m_t_cont}).  
In this situation, any slope larger than 1.5   
(i.e. $B>1.5$) requires a lower normalization  
$A$ consistent with what is generally observed.   
  
We show in Fig.~\ref{m_t} a comparison with   
the recent results from Finoguenov et al. (2001)   
and Allen et al. (2001). 
Both are consistent with our results.
For example, Allen et al. measure a normalization
that is 13 per cent lower than ours with a 
difference of about $0.4 \sigma$.    
  
Another way to consider the relation between the gas temperature and  
a physical quantity related to the overdensity typical of a cluster,  
is the $R - T$ relation, where $R$ is the radius of a sphere enclosing  
a given overdensity. It can be obtained   
directly from eqn.~\ref{eq:m-t}  
\begin{eqnarray}  
\lefteqn{ \frac{R}{100 h_{50}^{-1} \mbox{kpc}} =   
  5.07 \ \beta_T^{1/2} \ \left( \frac{50}{H_z}\right) \  
  \left( \frac{1000}{\Delta} \right)^{1/2} \  
  \left( \frac{T}{1 \mbox{keV}} \right)^{1/2}, }  \nonumber \\  
\lefteqn{ \log R_{100} = 0.71 +0.5 \log T \left(+ \log 
  \left(\beta_T^{1/2} f_{\Omega} \right) \right). }  
\label{eq:r-t}  
\end{eqnarray}  
  
\begin{figure}  
\epsfig{figure=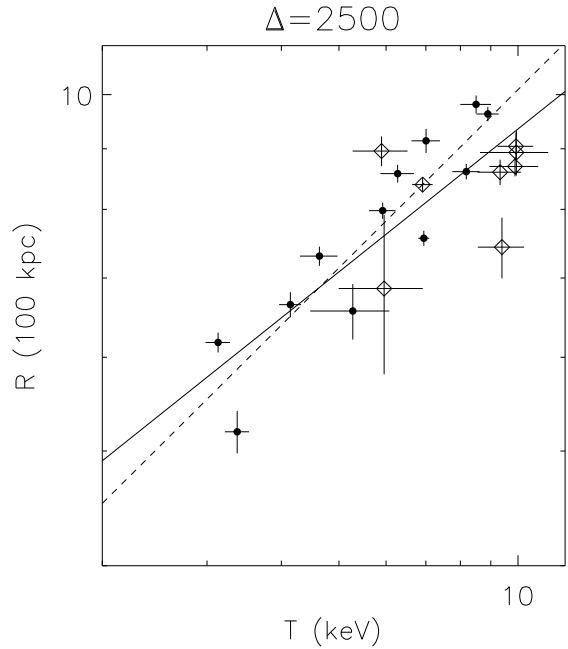,width=0.5\textwidth,angle=0}  
\caption{$R-T_{\rm mw}$ relation.  
The solid line represents the best-fit slope of $0.47 (\pm  
0.07)$; the dashed line is obtained fixing the slope to 1/2.  
{\it Filled circles} represent CF galaxy clusters, whereas  
{\it open squares} are NCF objects.  
} \label{r_t} \end{figure}

Fitting a power law as done above, we obtain   
$R_{100} = 3.27 (\pm 0.41) \times T_{\rm mw}^{0.47 (\pm 0.07)}$  
($\Delta =$ 2500; cf. Table~\ref{tab:fit} and Fig.~\ref{r_t}).  
Fixing the slope to 1/2 in accordance with what predicted from scaling laws  
the normalization is of $530 (\pm 79)$ kpc ($\Delta=1000$),  
only $0.4 \sigma$ below the value of 566 kpc measured in the  
hydrodynamics simulations of Evrard et al. (1996).  
  
\begin{figure}  
\epsfig{figure=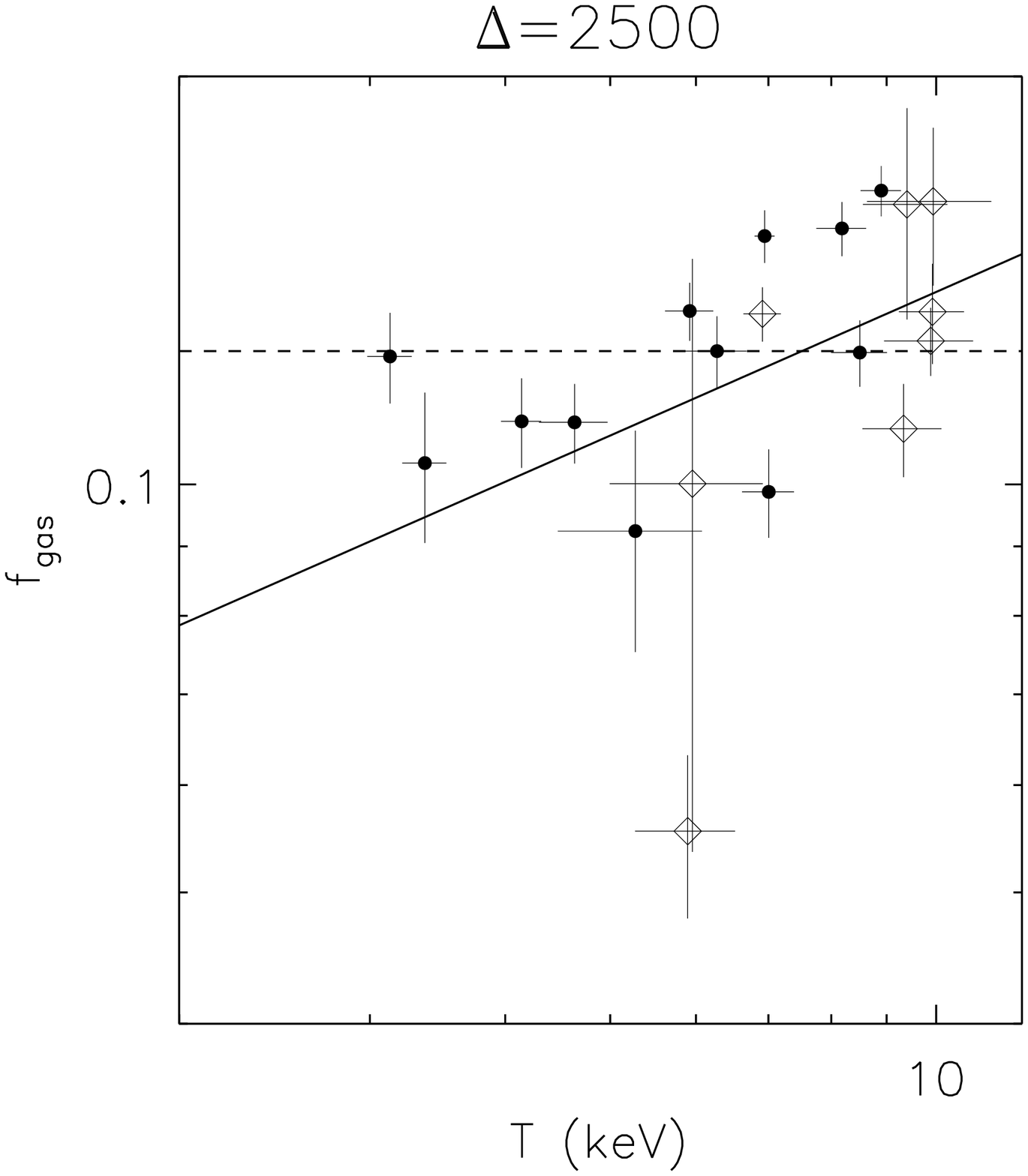,width=0.5\textwidth,angle=0}  
\caption{$f_{\rm gas}-T_{\rm mw}$ relation.  
The solid line represents the best-fit slope of $0.61 (\pm  
0.31)$; the dashed line is obtained fixing the slope to 0.  
{\it Filled circles} represent CF galaxy clusters, whereas  
{\it open squares} are NCF objects.  
} \label{fgas_t} \end{figure}  
  
\begin{figure*}  
\hbox{  
\epsfig{figure=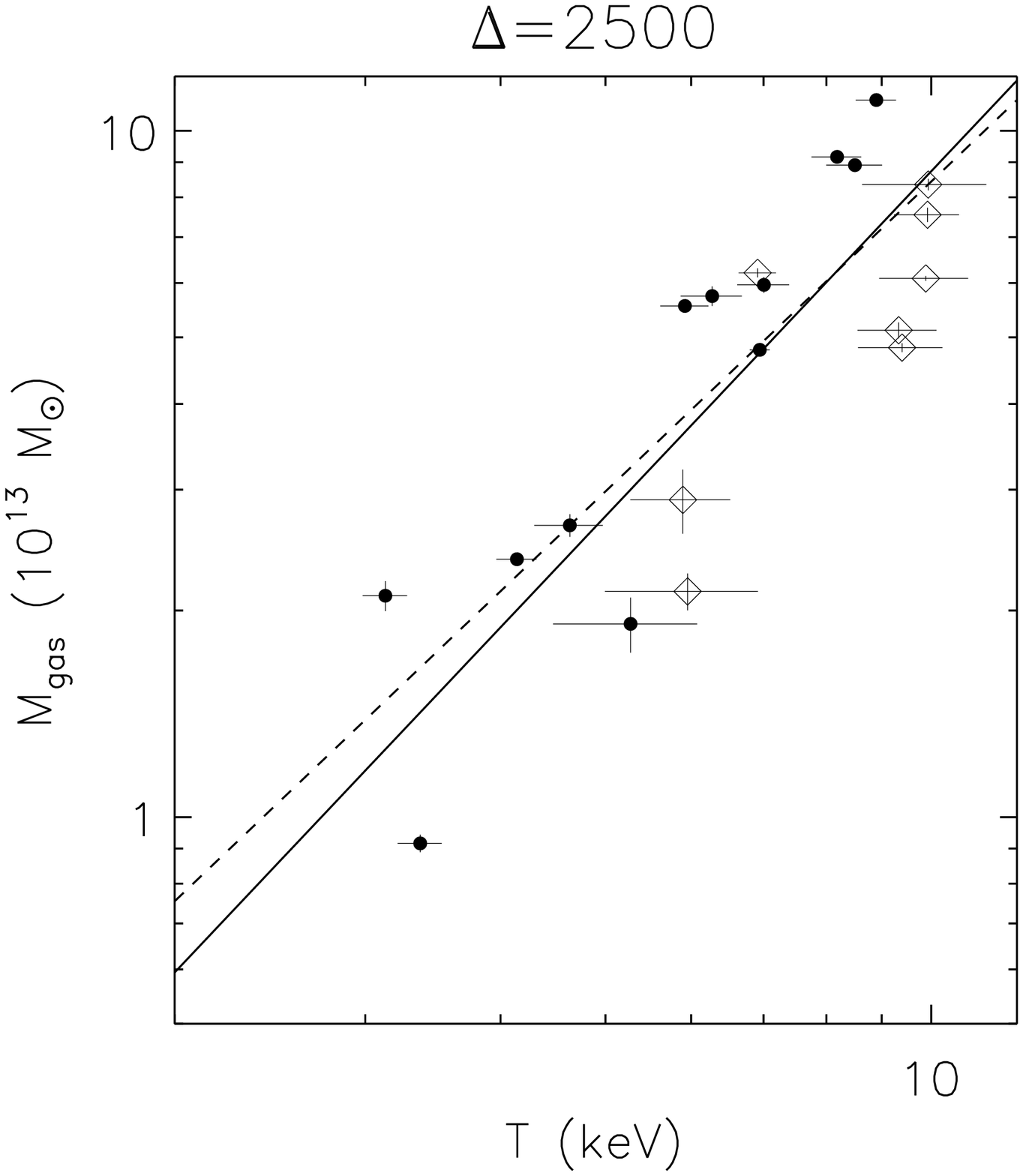,width=0.5\textwidth,angle=0}  
\epsfig{figure=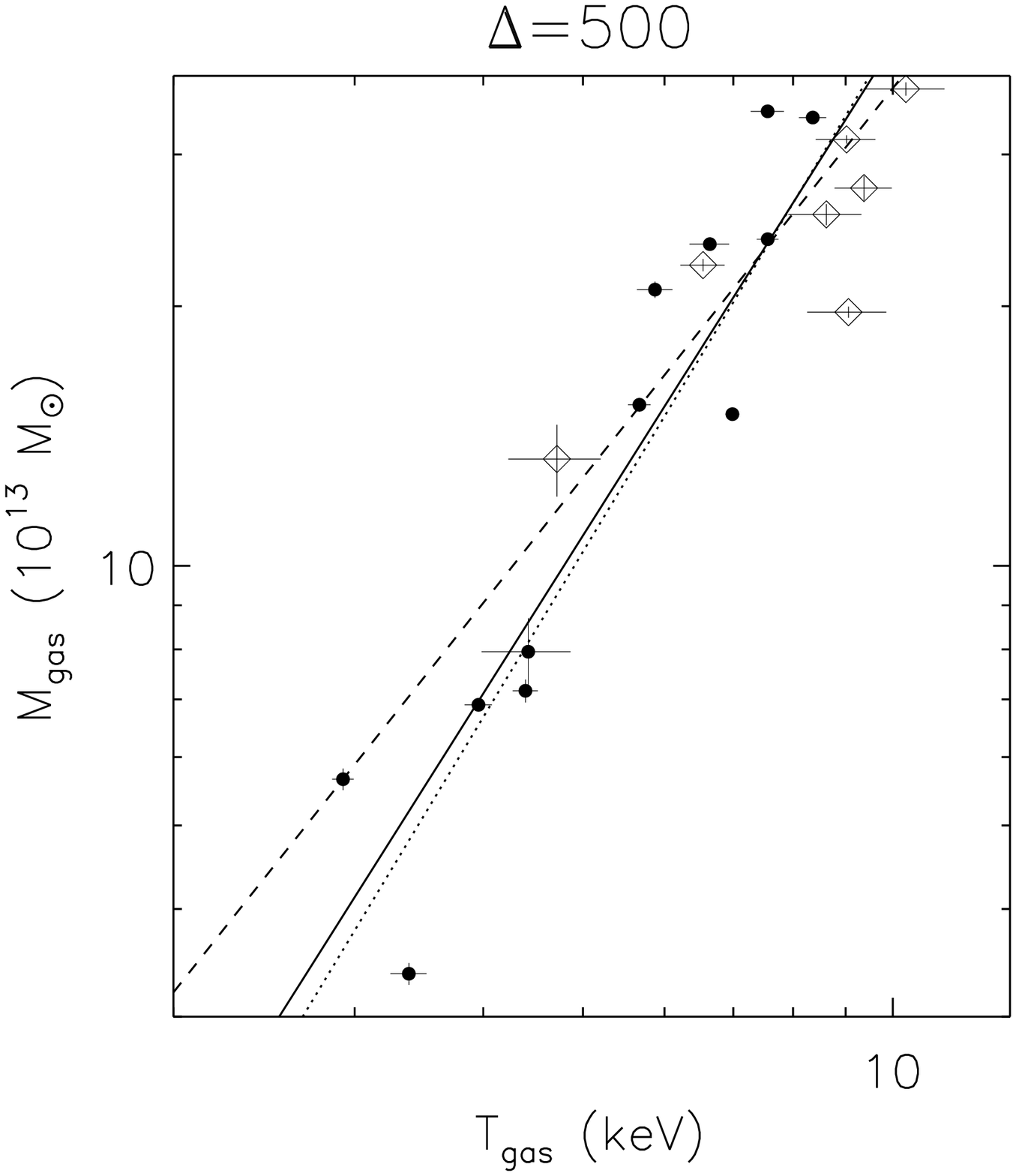,width=0.5\textwidth} }  
\caption{(Left panel) $M_{\rm gas}-T$ relation.   
The solid line represents the best-fit slope of $1.91 (\pm  
0.29)$; the dashed line is obtained fixing the slope to 3/2.  
{\it Filled circles} represent CF galaxy clusters, whereas  
{\it open squares} are NCF objects.  
(Right panel) $M_{\rm gas}-T_{\rm ew}$ relation.  
The solid line represents the best-fit $M_{\rm gas, 13} =0.52 
(\pm 0.19) \times T_{\rm ew}^{1.89 (\pm 0.20)}$;  
the dashed line is obtained fixing the slope to 3/2.  
The dotted line represents the best-fit results in Mohr et al. (1998).  
} \label{mgas_t} \end{figure*}  
  
\subsection{$M_{\rm gas} - T$ relation}

The gas mass fraction
\begin{equation}
f_{\rm gas}(r_{\Delta}) = \frac{ M_{\rm gas}(<r_{\Delta}) }
{ M_{\rm tot}(<r_{\Delta}) },
\label{eq:fgas}
\end{equation}
should be constant in a population of galaxy clusters that satisfies
the self-similar behaviour. Once the self-similarity is broken,
a dependence of $f_{\rm gas}$ on the temperature is expected
(Arnaud \& Evrard 1999, Viklinhin et al. 1999).
This propagates directly to the $M_{\rm gas} - T$ relation in the
following way: $M_{\rm gas} \approx f_{\rm gas} \ M_{\rm tot}
\propto T^{3/2 +\alpha}$, where a dependence of the form
$f_{\rm gas} \propto T^{\alpha}$ is introduced.
In Fig.~\ref{fgas_t} and Table~\ref{tab:fit}, 
we show that there is a slightly positive
correlation between $f_{\rm gas}$ and $T$,
with a slope larger than 0 by $1.9 \sigma$ 
both at $\Delta=$ 2500 and 1000.
As a consequence of this, the $M_{\rm gas} - T$ relation  
(Fig.~\ref{mgas_t}) tends to show a slope larger than 1.5.
  
Fitting with a power law the outer X-ray emission, Vikhlinin, Forman \&  
Jones (1999) found that  $M_{\rm gas} \propto T^{1.71 \pm 0.13}$  
(at the baryon overdensity of 1000 $\approx$ 200 in the dark matter  
overdensity).  
From the observed $L-T$ relation and applying X-ray scaling laws,  
Neumann \& Arnaud (2001) found that $M_{\rm gas} \propto T^{1.94}$  
is required.  
Mohr, Mathiesen \& Evrard (1999) estimated in a sample   
of nearby galaxy clusters that $M_{\rm gas} \propto T^{1.98 \pm 0.18}$   
within a density contrast of 500.   
At this overdensity, our sample includes 19 galaxy clusters, of which 
only two are directly observed (see Fig.\ref{rout_rd}).
If we extrapolate the physical quantities as described in Section~4,   
we obtain the correlation plotted in Fig.~\ref{mgas_t}   
(right panel) that shows   
an agreement in the overall trend with a slope of   
$1.89 (\pm 0.20)$.  
 
It it worth noting that, while at an overdensity of 2500 there
is a considerable segregation between CF and NCF systems,
when going out to $\Delta=$500 the segregation disappears.
This kind of behaviour is in line with expectations indeed,
while at $\Delta=2500$ the CF region contributes substantially
to the overall gas mass, at $\Delta=500$ the contribution is quite small.
The fact that segregation is actually not observed in our data at
an overdensity of 500 indicates that the extrapolations made to derive
quantities at $\Delta=500$ are reasonable ones.

\subsection{$L-T$ relation}  
  
The self-similar dependence between luminosity and temperature  
is written as   
\begin{eqnarray}  
L & \approx & \epsilon \ Vol   
 \approx \Lambda(T) \ n_{\rm gas}^2 \ R^3 \nonumber \\  
 & \approx & f_{\rm gas}^2 \ T^{1/2} \ M_{\rm tot}^2 \ R^{-3} \   
\approx H_z^2 \ \Delta \ f_{\rm gas}^2 \ T^{1/2} \ M_{\rm tot} \nonumber \\  
 & \approx &  H_z \ \Delta^{1/2} \ f_{\rm gas}^2 \ T^2,  
\label{eq:l-t0}  
\end{eqnarray}  
where $\epsilon$ is the X-ray emissivity and   
several, but reasonable, assumptions are made.   
Specifically: the cooling function, $\Lambda(T)$, is here described   
by only bremsstrahlung emission that is strictly valid only for  
$T >$ 2 keV; the gas fraction, $f_{\rm gas}$, and the radial   
dependence in the volume, $Vol \approx r_{\Delta}^3$, are not  
dependent on the temperature (see, e.g., Arnaud \& Evrard 1999).  
  
  
On the other hand, observations of the X-ray properties in   
cluster samples did show a departure from the predicted slope of 2   
with a measured value of about 3 (Mushotzky 1984, Edge \& Stewart 1991,   
David et al. 1993) and a scatter along the mean relation   
that can be reduced considering properly   
the effect of the cool cores (Fabian et al. 1994).  
Once the impact of the strength of cooling flows   
on the measured luminosity and emission-weighted temperature  
is taken into account, the $L-T$ correlation is shown to be tighter  
and more, but still not completely, consistent with the self-similar prediction  
(Allen \& Fabian 1998, Markevitch 1998, Arnaud \& Evrard 1999,  
Ettori, Allen \& Fabian 2001).  
However, a steeper $L-T$ dependence is expected when this relation  
is investigated over one order of magnitude in temperature, if  
the intracluster gas was pre-heated before the accretion in the  
potential well rising the entropy level in cooler systems   
(e.g. Ponman et al. 1996; Cavaliere, Menci \& Tozzi 1997;   
Ponman, Cannon \& Navarro 1999; Tozzi \& Norman 2001).   
Gas-dynamic simulations (e.g. Bialek, Evrard \& Mohr 2001,
Borgani et al. 2002) are capable of recovering 
the observed slope once the effect of
non-gravitational heat input from, e.g., AGNs and supernovae
in the order of about 100 keV cm$^2$, released either in an 
impulsive way or during the cluster formation history 
according to the star formation rate, is
taken into account.
  
\begin{figure}  
\epsfig{figure=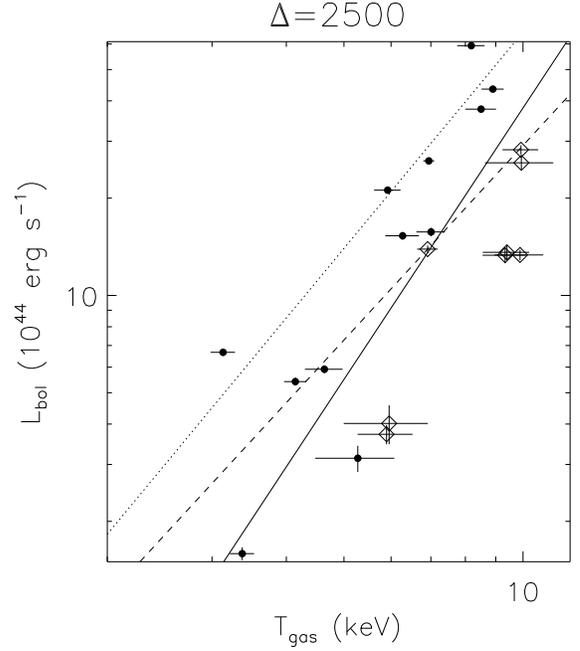,width=0.5\textwidth} 
\caption{$L-T_{\rm mw}$ relation.  
The solid line represents the best-fit slope of $2.79 (\pm  
0.55)$; the dashed line is obtained fixing the slope to 2.  
{\it Filled circles} represent CF galaxy clusters, whereas  
{\it open squares} are NCF objects.  
The dotted line shows the best-fit from Allen et al. (2001),
that follows the distribution of the CF clusters
in our sample.
} \label{l_t} \end{figure}  
  
In the temperature range investigated here, we observe   
a slope that is slightly higher, but still consistent within $2 \sigma$, 
with 2 using any of the three definitions  
of $T$ (see Table~\ref{tab:fit}) 
and at any overdensity (cf. Fig.~\ref{rel_dens}).  
 
\subsection{$L-M$ relation}  
  
From the observed correlation between total mass and gas temperature   
and between X-ray luminosity and temperature, it is straightforward  
to derive the dependence between gas luminosity and total mass.  
Combining eqn.~\ref{eq:m-t} and ~\ref{eq:l-t0}, we obtain  
that   
\begin{eqnarray}  
L & \approx & H_z^2 \ \Delta \ f_{\rm gas}^2 \ T^{1/2} \ M_{\rm tot} \nonumber \\ 
  & \approx & H_z^{7/3} \ \Delta^{7/6} \ f_{\rm gas}^2 \ M_{\rm tot}^{4/3}  
\label{eq:l-m0}  
\end{eqnarray}  
  
  
\begin{figure}  
\epsfig{figure=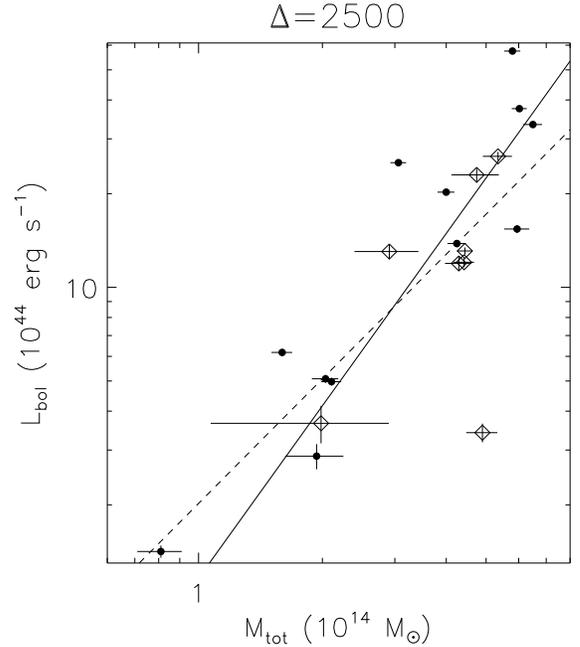,width=0.5\textwidth,angle=0}  
\caption{$L-M$ relation.   
The solid line represents the best-fit slope of $1.54 (\pm  
0.26)$; the dashed line is obtained fixing the slope to 4/3.  
{\it Filled circles} represent CF galaxy clusters, whereas  
{\it open squares} are NCF objects.  
} \label{l_m} \end{figure}  

Using a sample of 106 clusters observed with \rosat PSPC in the 
energy range 0.5--2.0 keV and with total masses estimated
through the $\beta$-model, Reiprich \& B\"ohringer (2002)
measure a slope of $1.80 (\pm 0.08)$ in the
$L_{\rm bol}-M_{200}$ relation.
This value is marginally consistent with our results    
enclosed between $1.84 (\pm 0.23)$, at $\Delta=$2500, and
$1.28 (\pm 0.15)$, at $\Delta=$1000
(cf. Table~\ref{tab:fit} and Fig.~\ref{rel_dens}).
These values lie between 4/3, as predicted 
in the self-similar scenario, and 11/6,
that is expected for a `minimum-entropy' scenario
in which cluster cores maintain
signature of an earlier galaxy formation activity
(cf. Evrard \& Henry 1991).

It is worth noticing that to use the X-ray galaxy clusters  
as cosmological probes and constraining  
the cluster mass function, the $L-T$ and $M-T$ relations  
are generally applied to recover from the observed  
X-ray luminosity a corresponding mass.  
For a given luminosity, $L$, this procedure carries an intrinsic scatter  
on the estimated mass, $\sigma_{\log M}$, that is the convolution of  
the scatter on the temperature originated in the $L-T$ relation,  
$\sigma_{\log T}$, with that present in the $M-T$ relation.  
From our estimates, this intrinsic scatter is about 0.20  
to be compared with the observed scatter in the $L-M$ relation  
of 0.14, that is about 30 per cent less 
(we have considered results for emission-weighted  
temperatures at the overdensity of 2500; these values 
are 0.24 and 0.20 at $\Delta=$1000, respectively).  
As a consequence of that, the direct application  
of the $L-M$ relation as tight as observed provides a more robust  
way to infer the cluster masses from large X-ray surveys  
(see also discussion on the cluster  
mass function estimated using the $L-M$ relation in   
Reiprich \& B\"ohringer 2002).  
  
\section{Summary and Conclusions}  
  
We have used resolved gas temperature and density profiles obtained   
from direct deprojection of \sax spectral data to measure the   
gravitating mass profiles of 20 nearby   
(0.010 $< z <$ 0.103; median redshift of 0.050) clusters.  
This sample is, to date, the largest for
which the physical quantities have all been obtained 
from the same dataset. 

At the overdensity (with respect to the critical density   
at a given redshift) of 2500, where 18 galaxy clusters   
in our sample have an observed temperature and luminosity,   
their gas mass weighted temperature spans over   
a factor of three in mass-weighted temperature
(3.1 keV $< T_{\rm X} <$ 9.9 keV; median value: 6.9 keV) and
two orders of magnitude in luminosity ($1.7 \times 10^{44}$ erg s$^{-1}$
$< L_{\rm X} < 6.1 \times 10^{45}$ erg s$^{-1}$; median value:
$1.5 \times 10^{45}$ erg s$^{-1}$).
We recover the total gravitating mass adopting either a 
King or a NFW functional form of the potential well.
We obtain that the X-ray mass estimates are generally in good
agreement with optical measurements.
After a selection that includes in the analyses
only systems with positive total mass
at the 95 per cent level of confidence for a given overdensity $\Delta$,
we investigate in a consistent and robust way several correlations
present between the quantities observed and
derived from the X-ray analysis.
Considering the high thermal energy associated with 
these objects, we are in the conditions to investigate the  
galaxy cluster scaling laws excluding systematics  
related to the energetic budget, but including  
statistical errors propagated from spatially-resolved  
gas density and temperature profiles.  

Considering two subsamples of clusters with (CF; 12 objects)   
and without (NCF; 8 obj.)
a mass deposition rate in the core (note that we do not  
consider any multi-phase gas in our spectral analysis, so that  
this classification only individuate systems more --CF--  
or less --NCF-- relaxed), we observe a segregation in the   
correlation planes according to this classification as originally  
suggested from Fabian et al. (1994) on the $L-T$ relation.  
Moreover, the scatter measured in these relations is reduced
when only CF systems are considered. 
These results point out an intrinsic scatter that is generally present  
in these correlations and is due to the dynamical status of the objects   
in exam. As expected, the scatter is greater for relations computed 
at larger overdensities, i.e. smaller radii, where the 
contribution from the more concentrated gas density is larger.
  
For each scaling relation that we investigate in the present work,   
we summarize here our main results:  
\begin{itemize}  
\item {\bf $M-T$ relation}: at any overdensity, the slope   
is consistent with a value of 1.5.
In particular, we obtain the best-fit robust relations  
$M_{14} = 0.20 (\pm 0.08) \times T_{\rm mw}^{1.54 (\pm 0.22)}$  
and $M_{14} = 0.25 (\pm 0.09) \times T_{\rm mw}^{1.5}$.  
at $\Delta = 2500$.  
A segregation between CF and NCF clusters is observable.  
When the slope is fixed to 1.5, the scatter is reduced  
by about 30 per cent for only CF objects ($\Delta = 2500$).
The normalization is lower than the results
of hydrodynamic simulations, but the large scatter observed 
makes large the uncertainties that we measure. 
In the same direction, the results on the $R-T$ relation show  
a slope in agreement with what is obtained
in numerical simulations and a normalization slightly lower
($R_{100} = 3.27 (\pm 0.41) \times T_{\rm mw}^{0.47 (\pm 0.07)}$  
and $R_{100} = 3.19 (\pm 0.32) \times T_{\rm mw}^{0.5}$).   
The evidence for a strong correlation between the fitted parameters  
in the $M-T$ relation
suggests that the physical interpretation of steeper slopes
and lower normalizations can be misleading
if the degeneracy  between these quantities is not properly
taken into account. 

\item {\bf $M_{\rm gas}-T$ relation}: both at $\Delta=$
2500 and 1000, we observe a marginally
significant trend between the gas mass fraction and the temperature  
(best-fit slope: $0.61 \pm 0.31$ at $\Delta=$2500).  
As a consequence of that, there is evidence, 
significant at less than $2 \sigma$, that the slope 
in the $M_{\rm gas}-T$ relation is different (and more steep) than  
the one present in the $M-T$ relation.  
At $\Delta = 2500$, we measure a slope of $1.91 (\pm 0.29)$.  
Due to the observed segregation between relaxed and not-relaxed  
clusters, the scatter is reduced by 20 per cent when 
CF objects are considered.   

\item {\bf $L-T$ relation}: our results are consistent with a slope of  
2 as predicted from the scaling law relations (maximal deviation 
of $2.1 \sigma$ at $\Delta=$1500).
At the overdensity of 2500,  we measure
$L_{44} = 0.06 (\pm 0.06) \times T_{\rm mw}^{2.79 (\pm 0.55)}$  
and $L_{44} = 0.29 (\pm 0.22) \times T_{\rm mw}^2$).  
The scatter is lowered by 20 per cent 
when only CF galaxy clusters are considered.
  
\item {\bf $L-M$ relation}: the central value measured on the 
slope of this relation varies between 
the self-similar and the `minimum-entropy' prediction.
The scatter in the relation is reduced by 25 per cent when  
only CF clusters are considered.  
We notice that the direct application of the $L-M$ relation   
as tightly as observed provides a more robust way   
(in other words, less scatter is propagated)   
to infer the cluster masses from large X-ray surveys  
than the combined application of the  
$L-T$ and $M-T$ relations.  
\end{itemize}  
  
In a forthcoming paper (Ettori et al., in preparation),
the cosmological implications of these relations and of
the gas mass fraction distribution observed at different
overdensity will be investigated.

\section*{Acknowledgements}
This paper has made use of linearized event files produced at the 
\sax Science Data Center.
We thank the referee, Dr. K. Mitsuda, for several suggestions
that helped us to improve the presentation of this work.
Jukka Nevalainen is thanked for useful comments.

\end{document}

%% file: m-t_tab1_bestfit.tex
\begin{tabular}{c@{\hspace{.7em}} c@{\hspace{.7em}} c@{\hspace{.7em}} r@{\hspace{.7em}}
c@{\hspace{.7em}} c@{\hspace{.7em}} c@{\hspace{.7em}} c@{\hspace{.7em}}
c@{\hspace{.7em}} c@{\hspace{.7em}} c@{\hspace{.7em}} c@{\hspace{.7em}}  }
Cluster & z & CF & $R_{\rm out}$ & & \multicolumn{3}{c}{King} & &  
  \multicolumn{3}{c}{NFW} \\
 & & & kpc/ \arcmin & & $r_{\rm s}$ & $c$ & $\chi^2$ (d.o.f.) & & 
  $r_{\rm s}$ & $c$ & $\chi^2$ (d.o.f.) \\
A85 & 0.0518 & y & 1323/ 16 & & 320 (31) & 6.74 (0.36) & 4.7 (4) & & 1282 (133) & 2.54 (0.23) & 4.3 (4) \\
A119 & 0.0440 & n & 1139/ 16 & & 584 (118) & 4.63 (0.48) & 7.6 (4) & & 1097 (48) & 2.66 (0.14) & 12.9 (4) \\
A426 (Perseus) & 0.0183 & y & 618/ 20 & & 102 (1) & 14.55 (0.13) & 55.9 (5) & & 392 (30) & 6.08 (0.27) & 25.4 (5) \\
A496 & 0.0320 & y & 845/ 16 & & 203 (14) & 8.25 (0.30) & 6.0 (4) & & 738 (66) & 3.37 (0.20) & 7.4 (4) \\
A754 & 0.0528 & n & 1683/ 20 & & 471 (91) & 5.56 (0.61) & 25.7 (5) & & 1619 (104) & 2.15 (0.13) & 18.3 (5) \\
A1367 & 0.0215 & n & 723/ 20 & & 718 (--) & 3.69 (--) & ... (5) & & 718 (--) & 2.68 (--) & ... (5) \\
A1656 (Coma) & 0.0232 & n & 777/ 20 & & 184 (48) & 10.06 (1.81) & 4.2 (5) & & 459 (242) & 5.42 (2.01) & 5.1 (5) \\
A1795 & 0.0632 & y & 1584/ 16 & & 314 (22) & 6.78 (0.28) & 1.5 (4) & & 1024 (218) & 2.93 (0.35) & 1.1 (4) \\
A2029 & 0.0767 & y & 1410/ 12 & & 427 (44) & 6.14 (0.31) & 9.5 (3) & & 1390 (127) & 2.61 (0.20) & 6.8 (3) \\
A2142 & 0.0899 & y & 2157/ 16 & & 477 (40) & 5.37 (0.26) & 2.6 (4) & & 1654 (285) & 2.16 (0.24) & 3.6 (4) \\
A2199 & 0.0309 & y & 1022/ 20 & & 175 (11) & 9.48 (0.35) & 3.7 (5) & & 560 (157) & 4.29 (0.69) & 4.0 (5) \\
A2256 & 0.0581 & n & 1469/ 16 & & 570 (68) & 4.57 (0.26) & 5.3 (4) & & 1422 (15) & 2.19 (0.04) & 20.0 (4) \\
A2319 & 0.0564 & n & 1430/ 16 & & 269 (101) & 7.65 (1.76) & 4.9 (4) & & 1301 (300) & 2.57 (0.68) & 3.9 (4) \\
A3266 & 0.0594 & n & 1873/ 20 & & 362 (74) & 6.28 (0.86) & 2.2 (5) & & 1576 (182) & 2.17 (0.21) & 1.9 (5) \\
A3376 & 0.0456 & n & 883/ 12 & & 105 (--) & 9.26 (--) & ... (3) & & 176 (--) & 6.78 (--) & ... (3) \\
A3526 (Centaurus) & 0.0104 & y & 356/ 20 & & 76 (7) & 15.47 (0.67) & 3.3 (5) & & 345 (48) & 5.82 (0.60) & 3.4 (5) \\
A3562 & 0.0483 & y & 1241/ 16 & & 197 (51) & 8.23 (1.49) & 5.8 (4) & & 340 (187) & 5.74 (2.50) & 6.8 (4) \\
A3571 & 0.0391 & y & 1275/ 20 & & 279 (30) & 8.11 (0.48) & 7.6 (5) & & 1122 (192) & 3.08 (0.40) & 6.2 (5) \\
A3627 & 0.0157 & n & 533/ 20 & & 188 (111) & 8.85 (2.55) & 18.9 (5) & & 517 (139) & 4.39 (1.57) & 19.4 (5) \\
2A0335 & 0.0349 & y & 917/ 16 & & 186 (12) & 8.29 (0.28) & 6.5 (4) & & 626 (143) & 3.61 (0.52) & 7.6 (4) \\
PKS0745 & 0.1028 & y & 1812/ 12 & & 400 (54) & 6.04 (0.42) & 5.6 (3) & & 1148 (174) & 2.87 (0.40) & 5.5 (3) \\
TRIANG & 0.0510 & n & 1631/ 20 & & 259 (39) & 8.61 (0.93) & 4.1 (5) & & 666 (255) & 4.47 (1.31) & 3.7 (5) \\
\end{tabular}

%% file: m-t_tab_mass.tex
\begin{tabular}{c@{\hspace{.7em}} c@{\hspace{.6em}} c@{\hspace{.7em}}
c@{\hspace{.7em}} c@{\hspace{.7em}} c@{\hspace{.7em}} c@{\hspace{.7em}}
c@{\hspace{.7em}} c@{\hspace{.7em}} } 
 Cluster & & $R_{\Delta}$ & $T(R_{\Delta})$ & $T_{\rm ew}$ & $T_{\rm mw}$ & $L_{\rm bol}$ & 
 $M_{\rm gas}$ & $M_{\rm tot}$ \\
 & & kpc & keV & keV & keV & $10^{44}$ erg s$^{-1}$ & $10^{13} M_{\odot}$ & 
   $10^{14} M_{\odot}$ \\ \\ 
 \multicolumn{9}{c}{ $\Delta=$2500 } \\ 
A85 & & 795 (14) & 5.81 (0.79) & 6.20 (0.26) & 6.28 (0.41) & 16.50 (0.45) & 5.32 (0.17) & 4.25 (0.22) \\ 
A119 & & 840 (24) & 5.04 (0.69) & 6.24 (0.64) & 5.90 (0.62) & 3.97 (0.27) & 2.72 (0.29) & 4.90 (0.42) \\ 
A426 & & 736 (11) & 7.23 (0.26) & 5.96 (0.09) & 6.94 (0.15) & 26.80 (0.28) & 4.67 (0.04) & 3.06 (0.13) \\ 
A496 & & 634 (16) & 3.86 (0.31) & 4.04 (0.12) & 4.14 (0.18) & 5.67 (0.16) & 2.27 (0.04) & 2.04 (0.15) \\ 
A754 & & 805 (15) & 9.66 (1.51) & 9.84 (0.75) & 9.89 (0.93) & 14.41 (0.53) & 5.64 (0.04) & 4.42 (0.25) \\ 
A1656 & & 720 (42) & 8.80 (0.98) & 9.65 (0.85) & 9.40 (0.84) & 14.04 (0.75) & 4.68 (0.07) & 2.91 (0.51) \\ 
A1795 & & 771 (12) & 5.62 (0.73) & 5.82 (0.14) & 5.92 (0.30) & 21.91 (0.36) & 5.37 (0.07) & 4.00 (0.19) \\ 
A2029 & & 895 (16) & 8.79 (1.00) & 7.83 (0.29) & 8.50 (0.50) & 41.33 (0.98) & 8.13 (0.16) & 6.50 (0.34) \\ 
A2142 & & 862 (12) & 8.66 (0.81) & 8.66 (0.26) & 8.90 (0.38) & 48.59 (0.97) & 9.92 (0.08) & 6.02 (0.25) \\ 
A2199 & & 642 (12) & 4.77 (0.85) & 4.46 (0.16) & 4.64 (0.34) & 6.73 (0.19) & 2.34 (0.09) & 2.10 (0.12) \\ 
A2256 & & 802 (12) & 6.41 (0.49) & 7.17 (0.24) & 6.91 (0.27) & 14.56 (0.38) & 5.93 (0.09) & 4.44 (0.19) \\ 
A2319 & & 821 (36) & 10.20 (2.64) & 9.78 (0.90) & 9.94 (1.30) & 28.00 (1.33) & 7.67 (0.15) & 4.74 (0.63) \\ 
A3266 & & 792 (20) & 9.08 (1.34) & 9.38 (0.67) & 9.33 (0.78) & 14.46 (0.77) & 4.72 (0.13) & 4.29 (0.32) \\ 
A3526 & & 476 (20) & 2.66 (0.19) & 3.43 (0.14) & 3.37 (0.16) & 1.73 (0.08) & 0.84 (0.02) & 0.81 (0.10) \\ 
A3562 & & 614 (33) & 4.86 (1.40) & 5.46 (0.59) & 5.28 (0.80) & 3.35 (0.31) & 1.79 (0.16) & 1.94 (0.31) \\ 
A3571 & & 900 (21) & 6.04 (0.61) & 7.40 (0.33) & 7.01 (0.39) & 15.97 (0.48) & 5.87 (0.15) & 5.95 (0.42) \\ 
A3627 & & 639 (98) & 5.39 (1.06) & 6.08 (0.98) & 5.96 (0.96) & 4.31 (0.59) & 1.99 (0.12) & 1.98 (0.91) \\ 
2A0335 & & 583 (11) & 2.94 (0.30) & 3.00 (0.07) & 3.13 (0.15) & 7.06 (0.18) & 1.98 (0.10) & 1.60 (0.09) \\ 
PKS0745 & & 841 (13) & 8.75 (1.14) & 7.43 (0.17) & 8.19 (0.43) & 60.67 (0.80) & 8.95 (0.07) & 5.79 (0.26) \\ 
TRIANG & & 859 (23) & 9.39 (1.34) & 10.21 (0.52) & 9.92 (0.69) & 29.72 (0.97) & 7.17 (0.17) & 5.34 (0.44) \\ 
 \\  \multicolumn{9}{c}{ $\Delta=$1000 } \\
A85 & & 1393 (31) & 4.05 (0.43) & 6.01 (0.24) & 5.40 (0.30) & 19.37 (0.49) & 11.33 (0.25) & 9.14 (0.61) \\ 
A119 & & 1467 (88) & 2.17 (0.30) & 5.25 (0.54) & 4.07 (0.43) & 6.61 (0.50) & 6.97 (0.67) & 10.44 (1.88) \\ 
A426 & & 1156 (22) & 7.48 (0.27) & 6.54 (0.10) & 7.61 (0.16) & 37.61 (0.39) & 9.09 (0.07) & 4.74 (0.27) \\ 
A496 & & 968 (29) & 3.31 (0.26) & 3.99 (0.12) & 3.88 (0.17) & 6.54 (0.19) & 4.24 (0.08) & 2.90 (0.26) \\ 
A754 & & 1449 (27) & 7.17 (0.86) & 9.39 (0.63) & 8.73 (0.66) & 23.96 (0.82) & 15.23 (0.17) & 10.32 (0.57) \\ 
A1656 & & 1078 (80) & 7.90 (0.88) & 9.30 (0.82) & 8.91 (0.80) & 22.20 (1.19) & 10.61 (0.16) & 3.90 (0.87) \\ 
A1795 & & 1321 (36) & 3.69 (0.44) & 5.73 (0.14) & 5.15 (0.25) & 23.91 (0.44) & 10.11 (0.19) & 8.05 (0.65) \\ 
A2029 & & 1561 (32) & 5.04 (0.57) & 7.67 (0.28) & 7.12 (0.42) & 46.63 (1.12) & 17.91 (0.27) & 13.80 (0.86) \\ 
A2142 & & 1426 (35) & 6.85 (0.72) & 8.49 (0.26) & 8.03 (0.38) & 58.18 (1.26) & 19.89 (0.38) & 10.90 (0.80) \\ 
A2199 & & 966 (22) & 3.85 (0.39) & 4.43 (0.12) & 4.41 (0.21) & 7.46 (0.18) & 4.09 (0.13) & 2.87 (0.19) \\ 
A2256 & & 1410 (46) & 3.91 (0.32) & 6.76 (0.33) & 5.71 (0.30) & 19.12 (0.72) & 13.50 (0.22) & 9.65 (0.95) \\ 
A2319 & & 1437 (86) & 11.39 (1.96) & 10.08 (0.91) & 10.60 (1.16) & 38.44 (1.87) & 17.89 (0.32) & 10.16 (1.83) \\ 
A3266 & & 1424 (44) & 6.66 (1.22) & 8.96 (0.73) & 8.21 (0.79) & 20.62 (1.14) & 13.01 (0.51) & 9.98 (0.93) \\ 
A3526 & & 694 (31) & 2.13 (0.15) & 3.41 (0.14) & 3.11 (0.15) & 2.67 (0.12) & 1.78 (0.05) & 1.00 (0.13) \\ 
A3562 & & 937 (67) & 3.22 (0.72) & 4.94 (0.49) & 4.20 (0.56) & 4.24 (0.39) & 3.86 (0.36) & 2.76 (0.59) \\ 
A3571 & & 1532 (53) & 2.24 (0.23) & 6.94 (0.31) & 5.20 (0.29) & 19.10 (0.53) & 13.18 (0.23) & 11.73 (1.21) \\ 
A3627 & & 968 (199) & 4.74 (0.94) & 5.59 (0.90) & 5.46 (0.88) & 10.29 (1.42) & 4.77 (0.29) & 2.76 (1.70) \\ 
2A0335 & & 890 (22) & 2.02 (0.20) & 2.96 (0.07) & 2.73 (0.13) & 7.61 (0.18) & 3.47 (0.10) & 2.27 (0.16) \\ 
PKS0745 & & 1446 (36) & 9.08 (1.22) & 7.51 (0.18) & 8.53 (0.48) & 65.11 (0.92) & 15.81 (0.21) & 11.78 (0.88) \\ 
TRIANG & & 1392 (67) & 6.67 (0.85) & 9.78 (0.62) & 8.69 (0.66) & 36.81 (1.45) & 15.17 (0.49) & 9.10 (1.32) \\ 
\end{tabular}

%% file: m-t_tab2.tex
\begin{tabular}{ l@{\hspace{.7em}} c@{\hspace{.7em}} c@{\hspace{.7em}} 
c@{\hspace{.7em}}  c c@{\hspace{.7em}} c@{\hspace{.7em}} c@{\hspace{.7em}} } \hline 
relation &  $\rho$ & Prob & $| n \sigma |$ & & $\rho$ & Prob & $| n \sigma |$ \\ 
 & \multicolumn{3}{c}{ $\Delta=$2500 } & & 
 \multicolumn{3}{c}{ $\Delta=$1000 } \\  \hline 
$M_{\Delta}-T_{\Delta}$ & 0.69 & 0.001 & 3.00 & & 0.54 & 0.017 & 2.29 \\
$R_{\Delta}-T_{\Delta}$ & 0.71 & $<$0.001 & 3.09 & & 0.54 & 0.017 & 2.29 \\
$L_{\Delta}-T_{\Delta}$ & 0.65 & 0.002 & 2.83 & & 0.74 & $<$0.001 & 3.16 \\
$M_{\rm gas, \Delta}-T_{\Delta}$ & 0.77 & $<$0.001 & 3.34 & & 0.79 & $<$0.001 & 3.35 \\
$f_{\rm gas, \Delta}-T_{\Delta}$ & 0.58 & 0.007 & 2.52 & & 0.31 & 0.204 & 1.30 \\
$L_{\Delta}-M_{\Delta}$ & 0.75 & $<$0.001 & 3.26 & & 0.64 & 0.003 & 2.70  \\
\end{tabular}

%% file: m-t_tab_fit.tex
\begin{tabular}{c@{\hspace{.8em}} c@{\hspace{.8em}} c@{\hspace{.7em}}
 c@{\hspace{.7em}} c@{\hspace{.7em}} c@{\hspace{.7em}} c@{\hspace{.7em}}
 c@{\hspace{.7em}} c@{\hspace{.7em}} }
 relation & & \multicolumn{3}{c}{ $\Delta=$2500 } & & 
 \multicolumn{3}{c}{ $\Delta=$1000 } \\
  & & $A$ & $B$ & $\sigma_{\log Y}$ & & $A$ & $B$ & $\sigma_{\log Y}$ \\ 
 $M_{14}-T_{\rm mw}$  & & -0.70 (0.18) & 1.54 (0.22) & 0.14& & -0.52 (0.28) & 1.76 (0.34) & 0.25 \\
  & & -0.60 (0.15) & 1.50 (fix) & 0.16& & -0.34 (0.20) & 1.50 (fix) & 0.24 \\
 $M_{14}-T_{\rm ew}$  & & -0.67 (0.20) & 1.51 (0.24) & 0.15& & -0.73 (0.27) & 1.94 (0.32) & 0.22 \\
  & & -0.60 (0.16) & 1.50 (fix) & 0.16& & -0.39 (0.18) & 1.50 (fix) & 0.20 \\
 $M_{14}-T(R)$  & & -0.61 (0.14) & 1.47 (0.18) & 0.15& & -0.52 (0.32) & 1.24 (0.45) & 0.31 \\
  & & -0.64 (0.16) & 1.50 (fix) & 0.15& & -0.24 (0.29) & 1.50 (fix) & 0.35 \\  \\
 $L_{44}-T_{\rm mw}$  & & -1.21 (0.47) & 2.79 (0.55) & 0.31& & -0.61 (0.26) & 2.37 (0.33) & 0.22 \\
  & & -0.54 (0.32) & 2.00 (fix) & 0.28& & -0.32 (0.24) & 2.00 (fix) & 0.21 \\
 $L_{44}-T_{\rm ew}$  & & -1.07 (0.47) & 2.64 (0.55) & 0.34& & -0.82 (0.34) & 2.54 (0.42) & 0.25 \\
  & & -0.54 (0.37) & 2.00 (fix) & 0.31& & -0.41 (0.27) & 2.00 (fix) & 0.23 \\
 $L_{44}-T(R)$  & & -1.09 (0.45) & 2.73 (0.54) & 0.29& & -0.61 (0.20) & 1.76 (0.27) & 0.27 \\
  & & -0.47 (0.33) & 2.00 (fix) & 0.26& & -0.13 (0.30) & 2.00 (fix) & 0.29 \\  \\ 
 $M_{\rm gas, 13}-T$  & & -0.93 (0.25) & 1.91 (0.29) & 0.16& & -0.35 (0.18) & 1.74 (0.22) & 0.16 \\
  & & -0.57 (0.18) & 1.50 (fix) & 0.15& & -0.20 (0.17) & 1.50 (fix) & 0.15 \\
 $R_{100}-T$  & & 0.51 (0.05) & 0.47 (0.07) & 0.04& & 0.65 (0.10) & 0.60 (0.13) & 0.08 \\
  & & 0.50 (0.04) & 0.50 (fix) & 0.05& & 0.72 (0.06) & 0.50 (fix) & 0.08 \\
$f_{\rm gas}-T$ & &-1.42 (0.27) & 0.61 (0.31) & 0.11& & -1.34 (0.26) & 0.66 (0.34) & 0.13 \\
  & & -0.90 (0.10) & 0.00 (fix) & 0.11& & -0.85 (0.08) & 0.00 (fix) & 0.12 \\
 $L_{44}-M_{14}$  & & 0.06 (0.15) & 1.84 (0.23) & 0.26& & 0.16 (0.14) & 1.28 (0.15) & 0.27 \\
  & & 0.31 (0.19) & 1.33 (fix) & 0.25& & 0.14 (0.21) & 1.33 (fix) & 0.27 \\
\end{tabular}